\documentclass[10pt,journal,compsoc,letterpaper]{IEEEtran}

\usepackage[utf8]{inputenc}
\usepackage[nocompress]{cite}
\usepackage{soul, color}
\usepackage{graphicx, lipsum}
\usepackage[caption=false,font=footnotesize,labelfont=sf,textfont=sf]{subfig}
\usepackage{float}
\usepackage[hidelinks]{hyperref}
\usepackage{booktabs}
\usepackage{adjustbox}
\usepackage{amsmath}
\usepackage{amssymb}
\usepackage{bm}
\usepackage{float}
\usepackage{graphicx}
\usepackage{array}
\usepackage{xspace}
\usepackage{fancyhdr}
\usepackage{filemod}
\usepackage{balance}

\usepackage{rotating}

\usepackage{stfloats}

\usepackage[dvipsnames]{xcolor}

\usepackage{tikz}
\usepackage{pgfplots}
\usepackage{pgfplotstable}
\pgfplotsset{compat=1.18}
\usetikzlibrary{patterns}
\usetikzlibrary{arrows.meta}
\usepgfplotslibrary{groupplots}

\newcommand{\mirage}{\textsc{Mirage}\xspace}

\definecolor{cpuColor}{HTML}{6C757D}
\definecolor{fftColor}{HTML}{0D6EFD}
\definecolor{tdColor}{HTML}{DC3545}
\definecolor{fft2Color}{HTML}{20C997}
\definecolor{fft4Color}{HTML}{6F42C1}
\definecolor{ueA}{HTML}{DC3545}
\definecolor{ueB}{HTML}{DA6974}
\definecolor{ueC}{HTML}{0D6EFD}
\definecolor{ueD}{HTML}{5B9BFC}
\definecolor{stageA}{HTML}{1565C0}
\definecolor{stageB}{HTML}{42A5F5}
\definecolor{stageC}{HTML}{FF9800}
\definecolor{stageD}{HTML}{EF6C00}
\definecolor{instA}{HTML}{1F77B4}
\definecolor{instB}{HTML}{FF7F0E}
\definecolor{instC}{HTML}{2CA02C}
\definecolor{instD}{HTML}{D62728}
\pgfplotsset{
    every axis/.append style={
        font=\footnotesize,
        tick label style={font=\scriptsize},
        label style={font=\footnotesize},
        legend style={font=\scriptsize},
        grid=both,
        grid style={dashed, gray!30},
        major grid style={dashed, gray!30},
        minor grid style={dotted, gray!20},
        line width=0.7pt,
        tick style={black, line width=0.5pt},
        axis line style={line width=0.5pt},
    },
    /pgf/number format/.cd, 1000 sep={\,},
}

\usepackage[acronyms,nonumberlist,nopostdot,nomain,nogroupskip,acronymlists={hidden}]{glossaries}
\newglossary[algh]{hidden}{acrh}{acnh}{Hidden Acronyms}
\glsdisablehyper
\newacronym{3gpp}{3GPP}{3rd Generation Partnership Project}
\newacronym{4g}{4G}{4th generation mobile network}
\newacronym{5g}{5G}{5th generation mobile network}
\newacronym{6g}{6G}{sixth-generation}
\newacronym{nextg}{NextG}{Next Generation}
\newacronym{5gc}{5GC}{5G Core}
\newacronym{adc}{ADC}{Analog to Digital Converter}
\newacronym{aerpaw}{AERPAW}{Aerial Experimentation and Research Platform for Advanced Wireless}
\newacronym{ai}{AI}{Artificial Intelligence}
\newacronym{aimd}{AIMD}{Additive Increase Multiplicative Decrease}
\newacronym{am}{AM}{Acknowledged Mode}
\newacronym{amc}{AMC}{Adaptive Modulation and Coding}
\newacronym{amf}{AMF}{Access and Mobility Management Function}
\newacronym{aops}{AOPS}{Adaptive Order Prediction Scheduling}
\newacronym{api}{API}{Application Programming Interface}
\newacronym{apn}{APN}{Access Point Name}
\newacronym{aqm}{AQM}{Active Queue Management}
\newacronym{ausf}{AUSF}{Authentication Server Function}
\newacronym{avc}{AVC}{Advanced Video Coding}
\newacronym{awgn}{AGWN}{Additive White Gaussian Noise}
\newacronym{balia}{BALIA}{Balanced Link Adaptation Algorithm}
\newacronym{bbu}{BBU}{Base Band Unit}
\newacronym{bdp}{BDP}{Bandwidth-Delay Product}
\newacronym{ber}{BER}{Bit Error Rate}
\newacronym{bf}{BF}{Beamforming}
\newacronym{bler}{BLER}{Block Error Rate}
\newacronym{brr}{BRR}{Bayesian Ridge Regressor}
\newacronym{bsr}{BSR}{Buffer Status Report}
\newacronym{bs}{BS}{Base Station}
\newacronym{bpsk}{BPSK}{Binary Phase-shift keying}
\newacronym{bss}{BSS}{Business Support System}
\newacronym{ca}{CA}{Carrier Aggregation}
\newacronym{caas}{CaaS}{Connectivity-as-a-Service}
\newacronym{cb}{CB}{Code Block}
\newacronym{cc}{CC}{Congestion Control}
\newacronym{ccid}{CCID}{Congestion Control ID}
\newacronym{cco}{CC}{Carrier Component}
\newacronym{cd}{CD}{Continuous Delivery}
\newacronym{cdd}{CDD}{Cyclic Delay Diversity}
\newacronym{cdf}{CDF}{Cumulative Distribution Function}
\newacronym{cdma}{CDMA}{Code-Division Multiple Access}
\newacronym{cdn}{CDN}{Content Distribution Network}
\newacronym{ci}{CI}{Continuous Integration}
\newacronym{cicd}{CI/CD}{Continuous Integration/Continuous Delivery}
\newacronym{cir}{CIR}{Channel Impulse Response}
\newacronym{cn}{CN}{Core Network}
\newacronym{codel}{CoDel}{Controlled Delay Management}
\newacronym{comac}{COMAC}{Converged Multi-Access and Core}
\newacronym{cord}{CORD}{Central Office Re-architected as a Datacenter}
\newacronym{cornet}{CORNET}{COgnitive Radio NETwork}
\newacronym{cosmos}{COSMOS}{Cloud Enhanced Open Software Defined Mobile Wireless Testbed for City-Scale Deployment}
\newacronym{cots}{COTS}{Commercial Off-the-Shelf}
\newacronym{cp}{CP}{Control Plane}
\newacronym{cpu}{CPU}{Central Processing Unit}
\newacronym{cqi}{CQI}{Channel Quality Information}
\newacronym{cr}{CR}{Cognitive Radio}
\newacronym{cran}{CRAN}{Cloud \gls{ran}}
\newacronym{crs}{CRS}{Cell Reference Signal}
\newacronym{csi}{CSI}{Channel State Information}
\newacronym{csirs}{CSI-RS}{Channel State Information - Reference Signal}
\newacronym{cu}{CU}{Central Unit}
\newacronym{d2tcp}{D$^2$TCP}{Deadline-aware Data center TCP}
\newacronym{d3}{D$^3$}{Deadline-Driven Delivery}
\newacronym{dac}{DAC}{Digital to Analog Converter}
\newacronym{dag}{DAG}{Directed Acyclic Graph}
\newacronym{darpa}{DARPA}{Defense Advanced Research Projects Agency}
\newacronym{das}{DAS}{Distributed Antenna System}
\newacronym{dash}{DASH}{Dynamic Adaptive Streaming over HTTP}
\newacronym{dbs}{DBS}{Deep Brain Stimulation}
\newacronym{dc}{DC}{Dual Connectivity}
\newacronym{dccp}{DCCP}{Datagram Congestion Control Protocol}
\newacronym{dce}{DCE}{Direct Code Execution}
\newacronym{dci}{DCI}{Downlink Control Information}
\newacronym{dcl}{DCL}{Dear Colleague Letter}
\newacronym{dctcp}{DCTCP}{Data Center TCP}
\newacronym{devops}{DevOps}{Development and Operations}
\newacronym{dl}{DL}{Deep Learning}
\newacronym{dma}{DMA}{Direct Memory Access}
\newacronym{dmr}{DMR}{Deadline Miss Ratio}
\newacronym{dmrs}{DMRS}{DeModulation Reference Signal}
\newacronym{drlcc}{DRL-CC}{Deep Reinforcement Learning Congestion Control}
\newacronym{drs}{DRS}{Discovery Reference Signal}
\newacronym{dt}{DT}{Digital Twin}
\newacronym{dtn}{DTN}{Digital Twin Network}
\newacronym{dtmn}{DTMN}{Digital Twin for Mobile Network}
\newacronym{dtwn}{DTWN}{Digital Twin Wireless Network}
\newacronym{du}{DU}{Distributed Unit}
\newacronym{e2e}{E2E}{end-to-end}
\newacronym{ecaas}{ECaaS}{Edge-Cloud-as-a-Service}
\newacronym{ecn}{ECN}{Explicit Congestion Notification}
\newacronym{edf}{EDF}{Earliest Deadline First}
\newacronym{em}{EM}{Electro-Magnetic}
\newacronym{embb}{eMBB}{Enhanced Mobile Broadband}
\newacronym{empower}{EMPOWER}{EMpowering transatlantic PlatfOrms for advanced WirEless Research}
\newacronym{enb}{eNB}{evolved Node Base}
\newacronym{endc}{EN-DC}{E-UTRAN-\gls{nr} \gls{dc}}
\newacronym{epc}{EPC}{Evolved Packet Core}
\newacronym{eps}{EPS}{Evolved Packet System}
\newacronym{es}{ES}{Edge Server}
\newacronym{etsi}{ETSI}{European Telecommunications Standards Institute}
\newacronym[firstplural=Estimated Times of Arrival (ETAs)]{eta}{ETA}{Estimated Time of Arrival}
\newacronym{eutran}{E-UTRAN}{Evolved Universal Terrestrial Access Network}
\newacronym{faas}{FaaS}{Function-as-a-Service}
\newacronym{fapi}{FAPI}{Functional Application Platform Interface}
\newacronym{fcc}{FCC}{Federal Communications Commission}
\newacronym{fdd}{FDD}{Frequency Division Duplexing}
\newacronym{fdm}{FDM}{Frequency Division Multiplexing}
\newacronym{fdma}{FDMA}{Frequency Division Multiple Access}
\newacronym{fed4fire}{FED4FIRE+}{Federation 4 Future Internet Research and Experimentation Plus}
\newacronym{fir}{FIR}{Finite Impulse Response}
\newacronym{fit}{FIT}{Future \acrlong{iot}}
\newacronym{fl}{FL}{Federated Learning}
\newacronym{fpga}{FPGA}{Field Programmable Gate Array}
\newacronym{fr2}{FR2}{Frequency Range 2}
\newacronym{fs}{FS}{Fast Switching}
\newacronym{fscc}{FSCC}{Flow Sharing Congestion Control}
\newacronym{ftp}{FTP}{File Transfer Protocol}
\newacronym{fw}{FW}{Flow Window}
\newacronym{ga128}{Ga}{Golay Sequence type A}
\newacronym{ge}{GE}{Gaussian Elimination}
\newacronym{glfsr}{GLFSR}{Galois Linear Feedback Shift Register}
\newacronym{gnb}{gNB}{Next Generation Node Base}
\newacronym{gold}{Gold}{Gold}
\newacronym{gop}{GOP}{Group of Pictures}
\newacronym{gpr}{GPR}{Gaussian Process Regressor}
\newacronym{gpu}{GPU}{Graphics Processing Unit}
\newacronym{gtp}{GTP}{GPRS Tunneling Protocol}
\newacronym{gtpc}{GTP-C}{GPRS Tunnelling Protocol Control Plane}
\newacronym{gtpu}{GTP-U}{GPRS Tunnelling Protocol User Plane}
\newacronym{gtpv2c}{GTPv2-C}{\gls{gtp} v2 - Control}
\newacronym{gw}{GW}{Gateway}
\newacronym{harq}{HARQ}{Hybrid Automatic Repeat reQuest}
\newacronym{hetnet}{HetNet}{Heterogeneous Network}
\newacronym{hh}{HH}{Hard Handover}
\newacronym{hol}{HOL}{Head-of-Line}
\newacronym{hqf}{HQF}{Highest-quality-first}
\newacronym{hss}{HSS}{Home Subscription Server}
\newacronym{http}{HTTP}{HyperText Transfer Protocol}
\newacronym{ia}{IA}{Initial Access}
\newacronym{iab}{IAB}{Integrated Access and Backhaul}
\newacronym{ic}{IC}{Incident Command}
\newacronym{ietf}{IETF}{Internet Engineering Task Force}
\newacronym{ifw}{IFW}{Interference Free Window}
\newacronym{imsi}{IMSI}{International Mobile Subscriber Identity}
\newacronym{imt}{IMT}{International Mobile Telecommunication}
\newacronym{iot}{IoT}{Internet of Things}
\newacronym{ip}{IP}{Internet Protocol}
\newacronym{iq}{IQ}{In-phase and Quadrature}
\newacronym{isi}{ISI}{Inter-Symbol Interference}
\newacronym{itu}{ITU}{International Telecommunication Union}
\newacronym{kpi}{KPI}{Key Performance Indicator}
\newacronym{kpm}{KPM}{Key Performance Measurement}
\newacronym{kvm}{KVM}{Kernel-based Virtual Machine}
\newacronym{lfsr}{LFSR}{Linear Feedback Shift Register}
\newacronym{los}{LoS}{line-of-sight}
\newacronym{ls}{LS}{Loosely Synchronised}
\newacronym{lsm}{LSM}{Link-to-System Mapping}
\newacronym{lstm}{LSTM}{Long Short Term Memory}
\newacronym{lte}{LTE}{Long Term Evolution}
\newacronym{lxc}{LXC}{Linux Container}
\newacronym{m2m}{M2M}{Machine to Machine}
\newacronym{mac}{MAC}{Medium Access Control}
\newacronym{mai}{MAI}{Multiple Access Interference}
\newacronym{manet}{MANET}{Mobile Ad Hoc Network}
\newacronym{mano}{MANO}{Management and Orchestration}
\newacronym{mc}{MC}{Multi-Connectivity}
\newacronym{mcc}{MCC}{Mobile Cloud Computing}
\newacronym{mchem}{MCHEM}{Massive Channel Emulator}
\newacronym{mcs}{MCS}{Modulation and Coding Scheme}
\newacronym{mec}{MEC}{Multi-access Edge Computing}
\newacronym{mec2}{MEC}{Mobile Edge Cloud}
\newacronym{mec3}{MEC}{Mobile Edge Computing}
\newacronym{mfc}{MFC}{Mobile Fog Computing}
\newacronym{mi}{MI}{Mutual Information}
\newacronym{mib}{MIB}{Master Information Block}
\newacronym{miesm}{MIESM}{Mutual Information Based Effective SINR}
\newacronym{mimo}{MIMO}{Multiple Input, Multiple Output}
\newacronym{mgen}{MGEN}{Multi-Generator}
\newacronym{ml}{ML}{Machine Learning}
\newacronym{mlr}{MLR}{Maximum-local-rate}
\newacronym[plural=\gls{mme}s,firstplural=Mobility Management Entities (MMEs)]{mme}{MME}{Mobility Management Entity}
\newacronym{mmtc}{mMTC}{Massive Machine-Type Communications}
\newacronym{mmwave}{mmWave}{millimeter wave}
\newacronym{mpdccp}{MP-DCCP}{Multipath Datagram Congestion Control Protocol}
\newacronym{mptcp}{MPTCP}{Multipath TCP}
\newacronym{mr}{MR}{Maximum Rate}
\newacronym{mrdc}{MR-DC}{Multi \gls{rat} \gls{dc}}
\newacronym{mse}{MSE}{Mean Square Error}
\newacronym{mss}{MSS}{Maximum Segment Size}
\newacronym{mt}{MT}{Mobile Termination}
\newacronym{mtd}{MTD}{Machine-Type Device}
\newacronym{mtu}{MTU}{Maximum Transmission Unit}
\newacronym{mumimo}{MU-MIMO}{Multi-user \gls{mimo}}
\newacronym{mvno}{MVNO}{Mobile Virtual Network Operator}
\newacronym{nalu}{NALU}{Network Abstraction Layer Unit}
\newacronym{nas}{NAS}{Network Attached Storage}
\newacronym{nbiot}{NB-IoT}{Narrow Band IoT}
\newacronym{nfv}{NFV}{Network Function Virtualization}
\newacronym{nfvi}{NFVI}{Network Function Virtualization Infrastructure}
\newacronym{nic}{NIC}{Network Interface Card}
\newacronym{nlos}{NLoS}{non-line-of-sight}
\newacronym{now}{NOW}{Non Overlapping Window}
\newacronym{nrdz}{NRDZ}{National Radio Dynamic Zone}
\newacronym{nsf}{NSF}{National Science Foundation}
\newacronym{nsm}{NSM}{Network Service Mesh}
\newacronym[type=hidden]{nr}{NR}{New Radio}
\newacronym{nrf}{NRF}{Network Repository Function}
\newacronym{nsa}{NSA}{Non Stand Alone}
\newacronym{nse}{NSE}{Network Slicing Engine}
\newacronym{nssf}{NSSF}{Network Slice Selection Function}
\newacronym{ntp}{NTP}{Network Time Protocol}
\newacronym{o2i}{O2I}{Outdoor to Indoor}
\newacronym{oai}{OAI}{OpenAirInterface}
\newacronym{oaicn}{OAI-CN}{\gls{oai} \acrlong{cn}}
\newacronym{oairan}{OAI-RAN}{\acrlong{oai} \acrlong{ran}}
\newacronym{oam}{OAM}{Operations, Administration and Maintenance}
\newacronym[plural=\gls{obu}s,firstplural=Onboard Units (OBUs)]{obu}{OBU}{Onboard Unit}
\newacronym{ofdm}{OFDM}{Orthogonal Frequency Division Multiplexing}
\newacronym{olia}{OLIA}{Opportunistic Linked Increase Algorithm}
\newacronym{omec}{OMEC}{Open Mobile Evolved Core}
\newacronym{onap}{ONAP}{Open Network Automation Platform}
\newacronym{onf}{ONF}{Open Networking Foundation}
\newacronym{onos}{ONOS}{Open Networking Operating System}
\newacronym{oom}{OOM}{\gls{onap} Operations Manager}
\newacronym{opnfv}{OPNFV}{Open Platform for \gls{nfv}}
\newacronym[type=hidden]{oran}{O-RAN}{Open RAN}
\newacronym{orbit}{ORBIT}{Open-Access Research Testbed for Next-Generation Wireless Networks}
\newacronym{os}{OS}{Operating System}
\newacronym{osc}{OSC}{O-RAN Software Community}
\newacronym{osm}{OSM}{Open Street Map}
\newacronym{oss}{OSS}{Operations Support System}
\newacronym{pa}{PA}{Position-aware}
\newacronym{pase}{PASE}{Prioritization, Arbitration, and Self-adjusting Endpoints}
\newacronym{pawr}{PAWR}{Platforms for Advanced Wireless Research}
\newacronym{pbch}{PBCH}{Physical Broadcast Channel}
\newacronym{pcef}{PCEF}{Policy and Charging Enforcement Function}
\newacronym{pcie}{PCIe}{Peripheral Component Interconnect Express}
\newacronym{pcfich}{PCFICH}{Physical Control Format Indicator Channel}
\newacronym{pcrf}{PCRF}{Policy and Charging Rules Function}
\newacronym{pdcch}{PDCCH}{Physical Downlink Control Channel}
\newacronym{pdcp}{PDCP}{Packet Data Convergence Protocol}
\newacronym{pdsch}{PDSCH}{Physical Downlink Shared Channel}
\newacronym{pdu}{PDU}{Packet Data Unit}
\newacronym{pdp}{PDP}{Power Delay Profile}
\newacronym{pf}{PF}{Proportional Fair}
\newacronym{pgw}{PGW}{Packet Gateway}
\newacronym{phich}{PHICH}{Physical Hybrid ARQ Indicator Channel}
\newacronym{phy}{PHY}{physical layer}
\newacronym{pl}{PL}{Path Loss}
\newacronym{pmch}{PMCH}{Physical Multicast Channel}
\newacronym{pmi}{PMI}{Precoding Matrix Indicators}
\newacronym{powder}{POWDER}{Platform for Open Wireless Data-driven Experimental Research}
\newacronym{ppo}{PPO}{Proximal Policy Optimization}
\newacronym{ppp}{PPP}{Poisson Point Process}
\newacronym{prach}{PRACH}{Physical Random Access Channel}
\newacronym{prb}{PRB}{Physical Resource Block}
\newacronym{psnr}{PSNR}{Peak Signal to Noise Ratio}
\newacronym{pss}{PSS}{Primary Synchronization Signal}
\newacronym{pucch}{PUCCH}{Physical Uplink Control Channel}
\newacronym{pusch}{PUSCH}{Physical Uplink Shared Channel}
\newacronym{qam}{QAM}{Quadrature Amplitude Modulation}
\newacronym{qci}{QCI}{\gls{qos} Class Identifier}
\newacronym{qoe}{QoE}{Quality of Experience}
\newacronym{qos}{QoS}{Quality of Service}
\newacronym{qtgui}{QT-GUI}{QT Graphical User Interface}
\newacronym{quic}{QUIC}{Quick UDP Internet Connections}
\newacronym{rach}{RACH}{Random Access Channel}
\newacronym{ran}{RAN}{Radio Access Network}
\newacronym{ra}{RA}{Random Access}
\newacronym{ack}{ACK}{Acknowledgment}
\newacronym[firstplural=Radio Access Technologies (RATs)]{rat}{RAT}{Radio Access Technology}
\newacronym{rc}{RC}{RAN Control}
\newacronym{rcn}{RCN}{Research Coordination Network}
\newacronym{rec}{REC}{Radio Edge Cloud}
\newacronym{red}{RED}{Random Early Detection}
\newacronym{renew}{RENEW}{Reconfigurable Eco-system for Next-generation End-to-end Wireless}
\newacronym{rf}{RF}{Radio Frequency}
\newacronym{rfc}{RFC}{Request for Comments}
\newacronym{rfr}{RFR}{Random Forest Regressor}
\newacronym{ric}{RIC}{RAN Intelligent Controller}
\newacronym{rlc}{RLC}{Radio Link Control}
\newacronym{rlf}{RLF}{Radio Link Failure}
\newacronym{rlnc}{RLNC}{Random Linear Network Coding}
\newacronym{rmr}{RMR}{RIC Message Router}
\newacronym{rmse}{RMSE}{Root Mean Squared Error}
\newacronym{rnis}{RNIS}{Radio Network Information Service}
\newacronym{rr}{RR}{Round Robin}
\newacronym{rrc}{RRC}{Radio Resource Control}
\newacronym{rrm}{RRM}{Radio Resource Management}
\newacronym{rru}{RRU}{Remote Radio Unit}
\newacronym{rs}{RS}{Remote Server}
\newacronym{rsrp}{RSRP}{Reference Signal Received Power}
\newacronym{rsrq}{RSRQ}{Reference Signal Received Quality}
\newacronym{rss}{RSS}{Received Signal Strength}
\newacronym{rssi}{RSSI}{Received Signal Strength Indicator}
\newacronym{rsu}{RSU}{Road-Side Unit}
\newacronym{rtt}{RTT}{Round Trip Time}
\newacronym{ru}{RU}{Radio Unit}
\newacronym{rw}{RW}{Receive Window}
\newacronym{rx}{RX}{Receiver}
\newacronym{s1ap}{S1AP}{S1 Application Protocol}
\newacronym{sa}{SA}{standalone}
\newacronym{sack}{SACK}{Selective Acknowledgment}
\newacronym{sap}{SAP}{Service Access Point}
\newacronym{sc2}{SC2}{Spectrum Collaboration Challenge}
\newacronym{scef}{SCEF}{Service Capability Exposure Function}
\newacronym{sch}{SCH}{Secondary Cell Handover}
\newacronym{scoot}{SCOOT}{Split Cycle Offset Optimization Technique}
\newacronym{sctp}{SCTP}{Stream Control Transmission Protocol}
\newacronym{sdap}{SDAP}{Service Data Adaptation Protocol}
\newacronym{shm}{SHM}{POSIX shared memory}
\newacronym{sd}{SD}{Slice Differentiator}
\newacronym{sdk}{SDK}{Software Development Kit}
\newacronym{sdm}{SDM}{Space Division Multiplexing}
\newacronym{sdma}{SDMA}{Spatial Division Multiple Access}
\newacronym{sdn}{SDN}{Software-defined Networking}
\newacronym{sdr}{SDR}{Software-defined Radio}
\newacronym{seba}{SEBA}{SDN-Enabled Broadband Access}
\newacronym{sgsn}{SGSN}{Serving GPRS Support Node}
\newacronym{sgw}{SGW}{Service Gateway}
\newacronym{si}{SI}{Study Item}
\newacronym{sib}{SIB}{Secondary Information Block}
\newacronym{sinr}{SINR}{Signal to Interference plus Noise Ratio}
\newacronym{sip}{SIP}{Session Initiation Protocol}
\newacronym{siso}{SISO}{Single Input, Single Output}
\newacronym{sla}{SLA}{Service Level Agreement}
\newacronym{sm}{SM}{Saturation Mode}
\newacronym{smf}{SMF}{Session Management Function}
\newacronym{smo}{SMO}{Service Management and Orchestration}
\newacronym{sms}{SMS}{Short Message Service}
\newacronym{smsgmsc}{SMS-GMSC}{\gls{sms}-Gateway}
\newacronym{snr}{SNR}{Signal-to-Noise-Ratio}
\newacronym{snssai}{S-NSSAI}{Single Network Slice Selection Assistance Information}
\newacronym{son}{SON}{Self-Organizing Network}
\newacronym{sptcp}{SPTCP}{Single Path TCP}
\newacronym{srb}{SRB}{Service Radio Bearer}
\newacronym{srn}{SRN}{Standard Radio Node}
\newacronym{srs}{SRS}{Sounding Reference Signal}
\newacronym{ss}{SS}{Synchronization Signal}
\newacronym{sss}{SSS}{Secondary Synchronization Signal}
\newacronym{st}{ST}{Spanning Tree}
\newacronym{svc}{SVC}{Scalable Video Coding}
\newacronym{t310}{T310}{Radio Link Monitoring Out-of-Sync Timer}
\newacronym{tb}{TB}{Transport Block}
\newacronym{tcp}{TCP}{Transmission Control Protocol}
\newacronym{tdd}{TDD}{Time Division Duplexing}
\newacronym{tdm}{TDM}{Time Division Multiplexing}
\newacronym{tdma}{TDMA}{Time Division Multiple Access}
\newacronym{tfl}{TfL}{Transport for London}
\newacronym{tfrc}{TFRC}{TCP-Friendly Rate Control}
\newacronym{tft}{TFT}{Traffic Flow Template}
\newacronym{tgen}{TGEN}{Traffic Generator}
\newacronym{tip}{TIP}{Telecom Infra Project}
\newacronym{tm}{TM}{Transparent Mode}
\newacronym{to}{TO}{Telco Operator}
\newacronym{toa}{ToA}{Time of Arrival}
\newacronym{tr}{TR}{Technical Report}
\newacronym{trp}{TRP}{Transmitter Receiver Pair}
\newacronym{ts}{TS}{Technical Specification}
\newacronym{tti}{TTI}{Transmission Time Interval}
\newacronym{ttt}{TTT}{Time-to-Trigger}
\newacronym{tx}{TX}{Transmitter}
\newacronym{uas}{UAS}{Unmanned Aerial System}
\newacronym{uav}{UAV}{Unmanned Aerial Vehicle}
\newacronym{udm}{UDM}{Unified Data Management}
\newacronym{udp}{UDP}{User Datagram Protocol}
\newacronym{udr}{UDR}{Unified Data Repository}
\newacronym{ue}{UE}{User Equipment}
\newacronym{uhd}{UHD}{\gls{usrp} Hardware Driver}
\newacronym{ul}{UL}{Uplink}
\newacronym{um}{UM}{Unacknowledged Mode}
\newacronym{uml}{UML}{Unified Modeling Language}
\newacronym{upa}{UPA}{Uniform Planar Array}
\newacronym{upf}{UPF}{User Plane Function}
\newacronym{urllc}{URLLC}{Ultra Reliable and Low Latency Communication}
\newacronym{usa}{U.S.}{United States}
\newacronym{usim}{USIM}{Universal Subscriber Identity Module}
\newacronym{usrp}{USRP}{Universal Software Radio Peripheral}
\newacronym{utc}{UTC}{Urban Traffic Control}
\newacronym{vim}{VIM}{Virtualization Infrastructure Manager}
\newacronym{vm}{VM}{Virtual Machine}
\newacronym{vnf}{VNF}{Virtual Network Function}
\newacronym{volte}{VoLTE}{Voice over \gls{lte}}
\newacronym{voltha}{VOLTHA}{Virtual OLT HArdware Abstraction}
\newacronym{vr}{VR}{Virtual Reality}
\newacronym{vran}{vRAN}{Virtualized \gls{ran}}
\newacronym{vss}{VSS}{Video Streaming Server}
\newacronym{wbf}{WBF}{Wired Bias Function}
\newacronym{wf}{WF}{Wired-first}
\newacronym{wi}{WI}{Wireless InSite}
\newacronym{wlan}{WLAN}{Wireless Local Area Network}
\newacronym{pnf}{PNF}{Physical Network Function}
\newacronym{drl}{DRL}{Deep Reinforcement Learning}
\newacronym{mtc}{MTC}{Machine-type Communications}
\newacronym{v2x}{V2X}{Vehicle-to-everything}
\newacronym{cast}{CaST}{Channel emulation scenario generator and Sounder Toolchain}
\newacronym{gui}{GUI}{Graphical User Interface}
\newacronym{ups}{UPS}{Uninterruptible Power Supply}
\newacronym{ota}{OTA}{Over-the-Air}
\newacronym{hitl}{HITL}{hardware-in-the-loop}
\newacronym{soc}{SoC}{System-on-Chip}
\newacronym{eeg}{EEG}{electroencephalogram}
\newacronym{ieeg}{iEEG}{intracranial electroencephalogram}
\newacronym{ecg}{ECG}{electrocardiogram}
\newacronym{fph}{FPH}{false positive per hour}
\newacronym{cnn}{CNN}{Convolutional Neural Network}
\newacronym{ban}{BAN}{Body Area Network}
\newacronym{roc}{ROC}{Receiver Operating Characteristic Curve}
\newacronym{auc}{AUC}{Area Under the Curve}
\newacronym{ecdf}{eCDF}{Empirical Cumulative Distribution Function}
\newacronym{deepchem}{DeepChEmM}{Deep Learning Channel Emulation Model}
\newacronym{svm}{SVM}{Support Vector Machine}
\newacronym{rbf}{RBF}{Radial Basis Function}
\newacronym{knn}{KNN}{$k$-Nearest Neighbors}
\newacronym{gbsm}{GBSM}{Geometry-Based Stochastic Model}
\newacronym{mlp}{MLP}{Multilayer Perceptron}
\newacronym{fc}{FC}{Fully Connected}
\newacronym{ae}{AE}{Autoencoder}
\newacronym{olla}{OLLA}{Outer Loop Link Adaptation}
\newacronym{evm}{EVM}{Error Vector Magnitude}
\newacronym{scs}{SCS}{subcarrier spacing}
\newacronym{aodt}{AODT}{Aerial Omniverse Digital Twin}
\newacronym{cfr}{CFR}{Channel Frequency Response}
\newacronym{hil}{HIL}{Hardware-in-the-loop}
\newacronym{fft}{FFT}{Fast Fourier Transform}
\newacronym{ocudu}{OCUDU}{Open RAN CU/DU}

\iftrue
\usepackage{tikzpagenodes,etoolbox}
\usetikzlibrary{calc}
\usepackage[contents={}]{background}
\AddEverypageHook{%
\ifnumequal{\thepage}{1}{%
    \tikz[remember picture,overlay]{%
        \node[
            draw,
            minimum width=0.9\textwidth,
            text width=0.9\textwidth,
            align=center,
            font=\footnotesize
        ]
        at ($(current page header area) - (0,-1pt)$)
        {%
            This work has been submitted to the IEEE for possible publication.
        };
    }%
}{}%end ifnumequal
}
\fi
\begin{document}

\title{
MIRAGE: A GPU-Accelerated IQ-Level Channel Emulator for Real-Time Wireless Digital Twins
%MIRAGE: A GPU-Accelerated Real-time IQ-Level Channel Emulator for Wireless Digital Twins
}

\author{Ali Saeizadeh, Angelo Feraudo, Avaneeth Anil, Leonardo Bonati, Pedram Johari, and Tommaso Melodia%
\IEEEcompsocitemizethanks{\IEEEcompsocthanksitem
The authors are with the Institute for Intelligent Networked Systems, Northeastern University, Boston, MA, U.S.A.
E-mail: \{saeizadeh.a, a.feraudo, lnu.avan, l.bonati, p.johari, melodia\}@northeastern.edu.
\IEEEcompsocthanksitem This work is supported by the National Telecommunications and Information Administration (NTIA)'s Public Wireless Supply Chain Innovation Fund (PWSCIF), Award No.~25-60-IF011.}}

\IEEEtitleabstractindextext{%
\begin{abstract}
Wireless digital twins enable network policies to be evaluated before deployment, but reproducing channel effects within a running protocol stack requires continuous waveform processing under strict timing constraints.
We present \mirage, a GPU-accelerated channel emulator that applies propagation effects directly to in-phase and quadrature samples exchanged by software 5G New Radio stacks.
\mirage combines batched multiuser, multiantenna filtering with timestamped shared-memory transport and radio-driver integrations for OpenAirInterface and OCUDU.
External propagation models supply channel responses asynchronously, while versioned buffers ensure consistent coefficients within each processing interval.
Experiments on one NVIDIA H100 GPU evaluate waveform accuracy, processing time, channel updates, and full-stack traffic delivery.
At 61.44 million samples per second, eight single-antenna links, four two-by-two links, and one four-by-four link meet the tested 0.5 ms bidirectional processing criterion.
Waveform errors remain within one integer quantization step per component, with no mixed coefficient versions across 16.7 million analyzed slots.
Integration with the X5G testbed enables slice resource allocations to be evaluated before deployment.
Both evaluated slices ultimately satisfy their throughput caps, requiring one physical reconfiguration under line-of-sight conditions and four under non-line-of-sight conditions.
These results demonstrate full-stack policy evaluation and expose prediction errors affecting transfer to a physical network.
\end{abstract}
}
\maketitle
\IEEEdisplaynontitleabstractindextext
\IEEEpeerreviewmaketitle

\section{Introduction}
\label{sec:intro}

%ALI    REVISED
Generative AI can reduce the effort required to translate ideas into network implementations and control policies~\cite{aghayev_genesis_2026}. As implementation becomes easier, verification and validation become increasingly important before deployment~\cite{deng2026swebenchpro}. In \glspl{ran}, testing must check more than software correctness. It must also assess protocol interactions, radio timing, and behavior under different traffic and channel conditions. Testing changes directly on a live network can affect users, while wireless variability complicates reproducible comparisons~\cite{sentosa2025cellreplay}. Wireless \glspl{dt} provide a controlled environment for assessing these changes before applying them to the physical network~\cite{alkhateeb2023dtvision,polese2023coloran}. Their value depends on reproducing the channel conditions and system behavior relevant to the experiment~\cite{VillaTRBJPM24}.

% The transition toward \gls{6g} networks and the maturing \gls{oran} ecosystem are reshaping how mobile networks are designed, validated, and operated. 
% Programmable, disaggregated \glspl{ran} controlled by \gls{ai}/\gls{ml}-driven xApps and rApps demand an experimentation paradigm in which new schedulers, slicing policies, and beam-management algorithms can be evaluated against realistic, reproducible, site-specific wireless conditions before deployment on production infrastructure~\cite{polese2023oran}. 
%Wireless \glspl{dt} have emerged as the leading candidate for this role: a virtual replica of the physical \gls{ran} that mirrors its state in real time and exposes the same control surfaces as the live deployment~\cite{alkhateeb2023dtvision}.

%
To support this evaluation, we focus on implementation changes and control policies whose effects depend on waveform reception and protocol behavior. We adopt three design requirements for the twin.
First, \emph{full-stack fidelity}: the twin should execute the protocol-stack components under evaluation and retain interactions involving HARQ retransmissions, link adaptation, RLC segmentation, and scheduling under contention.
Second, \emph{IQ-level channel application}: channel effects should be applied directly to the baseband \gls{iq} waveform, allowing multipath propagation and frequency-selective fading to affect signal reception and the resulting \gls{cqi}, \gls{bler}, and protocol behavior. Fidelity additionally requires validation of the channel model and its application to the waveform~\cite{VillaTRBJPM24,hofer2018real}.
Third, \emph{real-time operation and state exchange}: the twin should sustain the timing requirements of the radio stack while exchanging relevant state and control information with the physical \gls{ran}.

These requirements involve distinct timing constraints. IQ processing follows the configured sample rate and NR slot duration, which depends on the \gls{scs}~\cite{3gpp38211}. Sustaining this processing requires coordination among channel computation, buffering, and protocol-stack execution. Channel updates follow changes in the propagation environment, whereas monitoring and control exchanges follow the needs of the control task and can use O-RAN interfaces~\cite{polese2023oran}.

%Ali    REVISED: AIRMap added
Existing platforms address these requirements through different trade-offs. Hardware-based emulators support real-time, signal-level experimentation using dedicated infrastructure~\cite{polese2024colosseum,keysight_propsim,viavi_vertex}. Software platforms offer greater deployment flexibility, but their channel representation and execution model determine whether they exercise waveform reception and preserve radio timing~\cite{ryu2022emane,pawar2025chronos,oai_rfsimulator,iye2025owdt,mamaghani2026tinytwin,gurses2026achem}.
The design problem is to coordinate multi-link, multi-antenna channel computation with continuous stack execution while external channel responses change asynchronously. Batching reduces GPU dispatch overhead, but the processing path must also preserve sample ordering and use complete coefficient versions. Addressing these requirements together motivates the transport, computation, and update mechanisms developed here.
Section~\ref{sec:related} examines the closest approaches in detail.

%Ali    REVISED: MIMO ADDED
Real-time IQ-level channel emulation requires processing continuous sample streams within the timing budget of the radio stack.
For $K$ \glspl{ue}, $L$ channel taps, and $N_\mathrm{slot}$ samples per slot, direct time-domain convolution requires $\mathcal{O}(K n_\mathrm{tx} n_\mathrm{rx} L N_\mathrm{slot})$ operations per slot, where $n_\mathrm{tx}$ and $n_\mathrm{rx}$ denote the numbers of transmit and receive antennas per link.
The sample count per slot depends on the sampling rate $f_s$ and the slot timing associated with the configured \gls{scs}.
The processing cost therefore grows with the number of links, channel taps, and antenna pairs, making real-time execution increasingly demanding under multi-UE \gls{mimo} operation.
Recent GPU-based massive-MIMO systems show that execution granularity and orchestration affect processing latency and throughput~\cite{xie2025megastation}. For channel emulation, the operating range consequently depends on channel computation, data movement, and coordination with the radio stack; concurrent \gls{dt} instances add competition for shared resources.

\textbf{This work.} We present \mirage (\emph{Multi-instance IQ-level Real-time Accelerated GPU Emulator}), a GPU-accelerated channel emulator for full-stack wireless \glspl{dt}.
The core contribution is an integrated processing path that batches per-link SISO/MIMO filtering, exchanges timestamped IQ samples with software 5G NR stacks through stack-specific radio drivers, and applies complete channel-response versions while external models produce updates asynchronously.
As shown in Fig.~\ref{fig:system_arch}, a broker couples the resulting software replicas to a physical \gls{ran} through state and control exchanges.
This architecture supports evaluation of network changes under site-specific channel conditions before selected configurations are applied to the physical deployment.

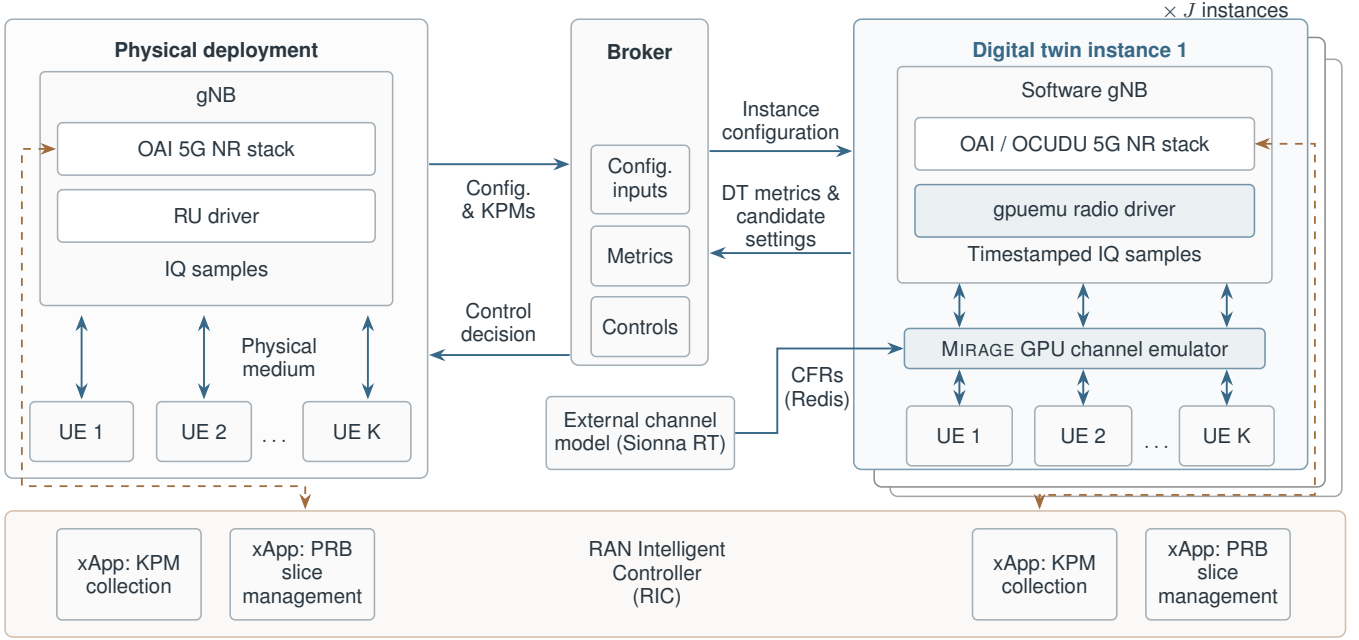
\begin{figure*}[t!]
    \centering
    \resizebox{\textwidth}{!}{%
        % Compile directly for a standalone preview, or input from the paper.
\ifdefined\pgfpicture
\else
  \documentclass{article}
  \usepackage[paperwidth=20cm,paperheight=10.5cm,margin=0.5cm]{geometry}
  \usepackage{tikz}
  \usetikzlibrary{arrows.meta}
  \pagestyle{empty}
  \def\architectureStandalone{1}
  \begin{document}
\fi
% Editable architecture figure. Requires tikz and arrows.meta.
\begingroup
\definecolor{archBlue}{HTML}{376887}
\definecolor{archOrange}{HTML}{A16D3D}
\definecolor{archInk}{HTML}{293640}
\definecolor{archRule}{HTML}{A4AFB7}
% Include inside a figure* using \resizebox{\textwidth}{!}{\input{...}}.
\begin{tikzpicture}[x=1cm,y=1cm,>=Stealth,
  font=\sffamily\footnotesize,text=archInk,
  panel/.style={draw=archRule,line width=.6pt,rounded corners=3pt,fill=white},
  stackbox/.style={draw=archRule,line width=.6pt,rounded corners=2pt,fill=black!2},
  layer/.style={draw=archRule,fill=white,line width=.6pt,rounded corners=2pt,
    minimum height=.72cm,align=center},
  driver/.style={draw=archBlue!70,fill=archBlue!10,line width=.6pt,rounded corners=2pt,
    minimum height=.72cm,align=center},
  ue/.style={draw=archRule,fill=black!2,line width=.6pt,rounded corners=2pt,
    minimum height=.82cm,align=center},
  brokerbox/.style={draw=archRule,fill=black!2,line width=.6pt,
    rounded corners=2pt,minimum width=1.35cm,align=center},
  flow/.style={->,draw=archBlue,line width=.75pt},
  radio/.style={<->,draw=archBlue,line width=.75pt},
  e2/.style={<->,draw=archOrange,dashed,line width=.7pt}]

% Parallel replicas, drawn behind the foreground instance.
\draw[panel,draw=black!35] (12.150,2.05) rectangle (18.350,8.05);
\draw[panel,draw=black!45] (11.930,2.18) rectangle (18.120,8.35);
\draw[panel,fill=black!1] (0.000,2.3) rectangle (5.800,8.6);
\draw[panel,fill=black!2] (7.770,3.85) rectangle (9.650,8.6);
\draw[panel,draw=archBlue!65,fill=archBlue!3] (11.650,2.42) rectangle (17.880,8.6);

\node[font=\sffamily\footnotesize\bfseries] at (2.900,8.15) {Physical deployment};
\node[font=\sffamily\footnotesize\bfseries] at (8.710,8.15) {Broker};
\node[font=\sffamily\footnotesize\bfseries,text=archBlue] at (14.760,8.15) {Digital twin instance 1};
\node[anchor=north east,text=archInk,font=\sffamily\footnotesize]
  at (17.740,8.96) {$\times\,J$ instances};

% Physical and emulated gNBs.
\draw[stackbox] (0.480,4.67) rectangle (5.320,7.88);
\node at (2.900,7.54) {gNB};
\node[layer,minimum width=4.36cm] at (2.900,6.82) {OAI 5G NR stack};
\node[layer,minimum width=4.36cm] at (2.900,5.90) {RU driver};
\node[font=\sffamily\footnotesize] at (2.900,5.15) {IQ samples};

\draw[stackbox] (12.250,4.98) rectangle (17.390,7.95);
\node at (14.820,7.61) {Software gNB};
\node[layer,minimum width=4.66cm] at (14.820,6.89) {OAI / OCUDU 5G NR stack};
\node[driver,minimum width=4.66cm] at (14.820,5.97) {gpuemu radio driver};
\node[text=archInk,font=\sffamily\footnotesize] at (14.820,5.35)
  {Timestamped IQ samples};

% Radio links and user equipment.
\foreach \x in {1.05,2.73,4.98}
  \draw[radio] (\x,4.54) -- (\x,3.38);
\node[align=center,font=\sffamily\footnotesize] at (3.760,3.96) {Physical\\medium};
\node[ue,minimum width=1.42cm] at (1.050,2.94) {UE 1};
\node[ue,minimum width=1.30cm] at (2.730,2.94) {UE 2};
\node[ue,minimum width=1.48cm] at (4.830,2.94) {UE K};
\node at (3.720,2.80) {$\cdots$};

\node[driver,minimum width=4.95cm,minimum height=.55cm,inner ysep=2pt]
  (mirage) at (14.820,4.08) {\textsc{Mirage} GPU channel emulator};
\node[ue,minimum width=1.45cm] at (13.100,2.88) {UE 1};
\node[ue,minimum width=1.33cm] at (14.800,2.88) {UE 2};
\node[ue,minimum width=1.30cm] at (16.770,2.88) {UE K};
\node at (15.830,2.73) {$\cdots$};
\foreach \x in {13.10,14.80,16.77} {
  \draw[radio] (\x,4.98) -- (\x,4.08 |- mirage.north);
  \draw[radio] (\x,4.08 |- mirage.south) -- (\x,3.29);
}

% Broker state, metrics, controls, and channel model.
\node[brokerbox,minimum height=.94cm] at (8.720,6.40) {Config.\\inputs};
\node[brokerbox,minimum height=.82cm] at (8.720,5.35) {Metrics};
\node[brokerbox,minimum height=.82cm] at (8.720,4.38) {Controls};
\node[brokerbox,minimum width=2.25cm,minimum height=1.0cm]
  (channelserver) at (8.720,2.92) {External channel\\model (Sionna RT)};
\draw[flow] (5.820,6.61) -- (7.750,6.61);
\node[align=center,font=\sffamily\footnotesize] at (6.776,6.13) {Config.\\\& KPMs};
\draw[flow] (7.750,3.99) -- (5.820,3.99);
\node[align=center,font=\sffamily\footnotesize] at (6.776,4.44) {Control\\decision};
\draw[flow] (9.670,6.79) -- (11.630,6.79);
\node[align=center,font=\sffamily\footnotesize] at (10.650,7.18) {Instance\\configuration};
\draw[flow] (11.610,5.39) -- (9.670,5.39);
\node[align=center,font=\sffamily\footnotesize] at (10.650,5.87) {DT metrics \&\\candidate\\settings};
\draw[flow] (channelserver.east) -- (10.600,2.92) -- (10.600,4.08) -- (mirage.west);
\node[align=center,font=\sffamily\footnotesize] at (11.150,3.53) {CFRs\\(Redis)};

% RIC and separate physical/DT control paths.
\draw[panel,draw=archOrange!45,fill=archOrange!4] (0.000,.12) rectangle (18.400,1.85);
\draw[e2] (0.700,6.82) -- (0.230,6.82) -- (0.230,2.19)
  -- (4.120,2.19) -- (4.120,1.87);
\draw[e2] (17.150,6.89) -- (17.980,6.89) -- (17.980,2.07)
  -- (14.200,2.07) -- (14.200,1.87);
\node[stackbox,minimum width=1.98cm,minimum height=1.25cm,align=center]
  at (1.700,.98) {xApp: KPM\\collection};
\node[stackbox,minimum width=1.98cm,minimum height=1.25cm,align=center]
  at (4.080,.98) {xApp: PRB\\slice\\management};
\node[align=center,font=\sffamily\footnotesize] at (8.950,.98)
  {RAN Intelligent\\Controller\\(RIC)};
\node[stackbox,minimum width=1.98cm,minimum height=1.25cm,align=center]
  at (14.270,.98) {xApp: KPM\\collection};
\node[stackbox,minimum width=1.98cm,minimum height=1.25cm,align=center]
  at (16.650,.98) {xApp: PRB\\slice\\management};
\end{tikzpicture}
\endgroup

\ifdefined\architectureStandalone
  \end{document}
\fi%
    }
    \caption{Wireless digital twin architecture with $J$ instances and $K$ UEs per instance. \mirage applies channel responses to timestamped IQ samples between software radio stacks. An external channel-model server delivers CFRs through Redis, independently of IQ processing. The broker coordinates configuration, measurements, and candidate settings; monitoring and control requests use the RIC/xApps and E2 paths shown by dashed arrows. Broker arrows denote logical information exchanges. The depicted monitoring and control integration is evaluated with OAI. The 5G cores are omitted for clarity.}
    \label{fig:system_arch}
\end{figure*}

%The channel-update path (Sionna~RT publishing $H_k(f)$ via Redis pub/sub) is decoupled from the signal-processing path (per-slot cuFFT pipeline) by a per-UE device buffer, so the slow ray-tracing cadence ($75$--$110$~ms per scene update) does not interfere with slot-rate IQ processing.

%
\subsection{Contributions}
\begin{itemize}
    \item \textbf{Batched GPU channel emulation integrated with a full 5G stack.}
    We design and implement a frequency-domain IQ-level channel
    emulator supporting per-\gls{ue} \gls{siso} and \gls{mimo}
    channels.
    Batched GPU processing and shared-memory IQ transport connect
    the emulator to \gls{oai} and OCUDU through radio-driver
    plugins, requiring no changes above the device-driver interface.
    OCUDU integration reuses the emulator and OAI UE source unchanged.
    We evaluate processing latency, signal fidelity, and end-to-end
    throughput under multi-UE and multi-antenna operation with OAI,
    and validate OCUDU interoperability in SISO operation.

    \item \textbf{Consistent asynchronous channel updates.}
    We separate external channel generation from IQ processing
    through a common \gls{cfr} interface and versioned GPU buffers.
    Each processing interval uses a complete coefficient version
    for each link and direction while subsequent responses are
    uploaded asynchronously. We validate update consistency and
    measure the delay before new responses are first used.

    \item \textbf{Concurrent full-stack instances and their operating limits.}
    We implement isolated \gls{dt} instances, each comprising a
    \gls{gnb}, \glspl{ue}, a GPU channel emulator, and a 5G core.
    We evaluate up to four concurrent instances sharing one NVIDIA
    H100 GPU and measure traffic delivery and radio-clock progression
    relative to wall time under a controlled offered load.
    The evaluation distinguishes successful traffic delivery from
    real-time execution and characterizes the effects of CPU
    allocation and concurrent execution on the shared host.

    \item \textbf{Closed-loop evaluation with a physical
    \gls{ran} testbed.}
    We integrate \mirage with the X5G testbed through a broker
    that exchanges network state and control information between
    the physical deployment and its digital twin.
    Using site-specific channels and corresponding radio
    configurations, we evaluate xApp-driven slice \gls{prb}
    allocation in the twin before applying selected quotas to
    the physical network.
    The experiments assess throughput agreement and control
    outcomes under \gls{los} and \gls{nlos} conditions, showing
    how prediction errors affect the need for subsequent
    physical reconfiguration.
\end{itemize}

% Declare the wide comparison table early for placement on page 3.
\begin{table*}[!t]
\centering
\caption{Architectural comparison of wireless emulation platforms and their cited implementations.}
\label{tab:related_work}
\renewcommand{\arraystretch}{1.12}
\setlength{\tabcolsep}{4pt}
\footnotesize
\begin{tabular}{@{}p{.21\textwidth}p{.39\textwidth}p{.35\textwidth}@{}}
\toprule
\textbf{Platform} & \textbf{Channel processing} & \textbf{Stack integration} \\
\midrule
Colosseum~\cite{polese2024colosseum}
 & FPGA signal-level emulation & SDR-based radio nodes \\
AODT~\cite{nvidia_aodt}
 & Scene-based propagation; GPU simulation & Integrated RAN simulation environment \\
5G-EMANE~\cite{ryu2022emane}
 & Abstracted PHY behavior & Cellular stack with EMANE \\
Chronos~\cite{pawar2025chronos}
 & Abstracted PHY; virtualized time & FAPI MAC--PHY interface \\
OAI \texttt{rfsim}~\cite{oai_rfsimulator}
 & Optional channel models; CPU-paced execution & TCP radio-sample interface \\
OAI \texttt{vrtsim}~\cite{oai_vrtsim}
 & Built-in models or external CIRs; adjustable pacing & Shared-memory interface; one gNB/UE pair \\
OWDT~\cite{iye2025owdt}
 & Sionna RT channels applied through OAI & OAI stack with Sionna RT \\
Tiny-Twin~\cite{mamaghani2026tinytwin}
 & CPU time-varying multi-tap filtering & Full 5G stack with per-UE channels \\
Sionna Research Kit~\cite{nvidia_srk_channel_2026}
 & CUDA SISO filtering; file or streamed CIRs & Channel plugin within OAI gNB \\
ACHEM~\cite{gurses2026achem}
 & IQ impairments; MIMO and mobility & Virtual USRP interface for UHD applications \\
\midrule
\mirage
 & Batched GPU SISO/MIMO filtering; versioned CFR updates & OAI and OCUDU radio drivers; timestamped shared-memory IQ \\
\bottomrule
\end{tabular}
\end{table*}

\section{Related Work}
\label{sec:related}

We review wireless emulation platforms, channel-modeling methods, and mechanisms for coupling digital twins to physical networks. Table~\ref{tab:related_work} summarizes the platform architectures.

\subsection{Hardware-Based and Integrated Platforms}

Colosseum~\cite{polese2024colosseum} combines \gls{usrp} nodes with an \gls{fpga} channel emulator, while commercial systems~\cite{keysight_propsim,viavi_vertex} provide dedicated channel-emulation appliances. Xie~et~al.~\cite{xie2025mimoemulator} present an 80-channel bidirectional transceiver and baseband platform for massive-MIMO emulation. These systems support signal-level experiments but require dedicated radio and emulation hardware. NVIDIA's \gls{aodt}~\cite{nvidia_aodt} instead combines scene-based propagation with GPU-accelerated RAN simulation. Our work targets a software channel emulator connected to an existing protocol stack through its radio-driver interface.

\subsection{Platforms with Abstracted PHY Processing}

5G-EMANE~\cite{ryu2022emane} and Chronos~\cite{pawar2025chronos} reduce computation by abstracting waveform-level \gls{phy} processing. Chronos combines a \gls{fapi} MAC--PHY interface~\cite{scf_fapi} with virtualized time for cloud-based experiments. Such abstractions support larger simulated networks, but do not exercise receiver processing on channel-impaired IQ samples, which is central to our evaluation target.

\subsection{Full-Stack IQ-Level Emulation}

The \gls{oai} \texttt{rfsim} replaces the radio driver with TCP-based IQ exchange and optional channel modeling; its execution can run faster or slower than wall-clock time~\cite{oai_rfsimulator}. VRTSIM uses shared-memory IQ exchange with a configurable timescale and supports built-in channel models and external CIRs through its \texttt{taps\_client} interface~\cite{oai_vrtsim}. The cited VRTSIM implementation documents a one-gNB/one-UE limit. Thus, shared-memory transport and external channel injection are existing capabilities; our contribution concerns their integration with batched multi-link GPU processing and coefficient-consistency mechanisms. The srsRAN 4G ZeroMQ interface similarly provides radio-sample transport that can be combined with external processing~\cite{srsran_zmq}.

The Open Wireless Digital Twin (OWDT)~\cite{iye2025owdt} couples OAI with Sionna RT for site-specific mobility emulation. Tiny-Twin~\cite{mamaghani2026tinytwin} applies time-varying, multi-tap channels on commodity CPUs, with per-UE isolation and RIC integration. ACHEM~\cite{gurses2026achem} virtualizes the USRP interface for UHD-based applications, including GNU Radio, srsRAN, and OAI, and supports channel impairments, mobility, and MIMO. These approaches provide full-stack channel emulation through different computation and radio-interface designs.

The Sionna Research Kit provides a particularly close GPU-based reference: its CUDA plugin applies CIRs within the OAI gNB in USRP and RF-simulator configurations~\cite{nvidia_srk_channel_2026}. The cited implementation supports SISO and accepts file-based or streamed CIRs. \mirage combines batched multi-link SISO/MIMO filtering with timestamped shared-memory transport and per-interval selection of complete channel-response versions. We evaluate this combination through waveform and update-consistency tests, bidirectional timing measurements, and full-stack delivery experiments, including concurrent instances.

\subsection{Channel Fidelity and Live Coupling}

Channel-model accuracy and emulator correctness require distinct validation. Hofer~et~al.~\cite{hofer2018real} compare simulated and emulated time-varying channel responses, while Villa~et~al.~\cite{VillaTRBJPM24} validate Colosseum scenarios against physical experiments. Rauf~et~al.~\cite{rauf2026kpi} examine system-level KPI deviations between Sionna RT and measured channels in an OAI hardware-in-the-loop testbed. These studies inform the separation of waveform validation from physical-network agreement in our evaluation.

Pegurri~et~al.~\cite{pegurri2025sionnans3} integrate Sionna RT with the event-driven ns-3 simulator, and AIRMap~\cite{saeizadeh2025airmap} estimates site-specific radio maps. The former evaluates channel effects through network-layer and application-layer simulation, while the latter supplies propagation estimates; neither provides the continuous IQ-processing path considered here. TwiNet~\cite{robinson_twinet_2024} addresses live coupling through bidirectional MQTT exchange between physical networks and digital replicas. Our testbed study connects IQ-level emulation to such a state-and-control loop and examines how prediction errors affect physical reconfiguration.

\section{System Architecture}
\label{sec:design}

Figure~\ref{fig:system_arch} shows the integration of \mirage into a full-stack wireless \gls{dt}. Each twin runs a software \gls{gnb}, software \glspl{ue}, and a 5G core, with \mirage applying per-link channel responses to the \gls{iq} samples exchanged between the radio stacks. The architecture separates continuous IQ processing from asynchronous channel generation and network control.

\subsection{Overview}

Each instance contains an OAI or OCUDU gNB, OAI software UEs, an Open5GS core, and the \mirage channel emulator. Stack-specific radio drivers adapt the gNB interface to a common IQ transport; channel processing is independent of the selected gNB stack. The monitoring and control integration and physical-network case study described here use OAI. Three paths connect these components to the channel-model server and the control system. The \emph{IQ path} carries timestamped baseband samples between the gNB, emulator, and UEs through shared memory. The \emph{channel-update path} supplies externally generated per-link responses to the emulator through Redis. The \emph{monitoring and control path} carries gNB measurements and resource-allocation requests through the E2 interface, with the broker coordinating information exchange and experiment execution. Only the IQ path follows the radio stack's sample-processing deadlines. Section~\ref{sec:sim} describes its transport and GPU processing mechanisms.

The physical deployment used for validation is the X5G testbed at Northeastern University~\cite{villa2025x5g}, with an OAI gNB, commercial Foxconn \glspl{ru}, and commercial handsets. The twin uses corresponding gNB radio settings and replaces the handsets with OAI software UEs. The physical network and the twin use matching radio settings, but their receivers can behave differently. Section~\ref{subsec:dt_setup} specifies the experimental configuration.

Independent instances support concurrent experiments with different candidate configurations. Their execution rate depends on the available CPU and GPU resources. Section~\ref{subsec:eval_multiverse} evaluates both traffic delivery and radio-clock progression under concurrent operation; Section~\ref{sec:dt} evaluates an iterative slice-control procedure using one twin and the physical deployment.

\subsection{Broker}
\label{subsec:broker}

The broker is a Python service that provisions containerized DT instances and coordinates configuration exchange, measurement access, and candidate-policy evaluation. Redis provides storage and messaging for these exchanges. The broker uses scene descriptors, UE locations, and radio settings to configure new instances. Each instance starts its own protocol processes; live HARQ state and packet buffers are not transferred from the physical network.

In the physical-network case study, the environmental inputs are an imported CAD scene and transmitter and receiver locations measured against the floor plan. Radio settings are matched through experiment configuration. During operation, the monitoring xApp writes per-UE throughput measurements to Redis, and the channel-model server publishes CFR updates. This experiment uses a configured scene rather than continuous sensing of the physical environment.

The control logic uses the measurements to assess candidate PRB quotas in the twin. The broker coordinates delivery of selected settings to the control xApp, which issues the corresponding request to the target gNB through E2. Thus, the broker manages experiment execution and information exchange, while the xApps provide the monitoring and control interfaces.

The physical deployment and the twin exchange configuration and measurements without IQ-sample alignment. The broker remains outside the slot-rate IQ path, and each instance maintains its own radio timing. Channel updates and control exchanges proceed asynchronously.

\subsection{Channel Generation and Delivery}
\label{subsec:channel_model_overview}

An external channel-model server supplies per-link \glspl{cfr}. Our reference deployment uses Sionna RT~\cite{hoydis_sionna_2023} to compute site-specific propagation from scene geometry, material properties, and transmitter and receiver locations. The emulator accepts scalar responses for \gls{siso} links and matrix-valued responses for \gls{mimo} links.

The server publishes responses through Redis to the emulator's channel subscriber. At the start of each processing interval, each direction selects the latest accepted response for each link and retains that version throughout the interval. Channel generation determines how frequently propagation coefficients are refreshed, while the IQ path continues processing with the available coefficients. Section~\ref{subsec:channel_model} describes version consistency and the frequency-grid, antenna-layout, and data-format requirements for alternative modeling backends.

Receiver noise is configured separately from the propagation response, as described in Section~\ref{subsec:multi_ue}.

\subsection{Near-RT RIC and Control xApps}

We use the OSC Near-RT \gls{ric}~\cite{osc_nearrtric} with monitoring and control xApps built using xDevSM~\cite{feraudo2026xdevsm}.\footnote{\url{https://github.com/wineslab/xDevSM}}
The monitoring xApp collects per-UE measurements through the E2 \gls{kpm} service model~\cite{oran_e2sm_kpm} and stores them in Redis for policy evaluation and comparison between the physical and virtual domains.

The control xApp uses the slice-level PRB-quota action of the E2 \gls{rc} service model~\cite{oran_e2sm_rc} to configure resource allocation on the target gNB. Candidate quotas are evaluated on the DT-side gNB before selected values are applied to the physical gNB. Subsequent monitoring assesses the physical response and supports further evaluation when a selected quota does not meet the policy objective. Section~\ref{subsec:dt_loop} details the procedure and Section~\ref{subsec:cl_results} reports its outcomes.

\section{CUDA-Based Channel Emulation}
\label{sec:sim}

\mirage applies per-link channel responses to timestamped \gls{iq} streams through GPU filtering, shared-memory transport, and asynchronous coefficient updates. We describe the single-gNB path used in the evaluation. The implementation sets configuration limits of 64 UEs and four transmit and receive antennas per link; Section~\ref{sec:eval} reports the evaluated configurations and their timing limits.

\subsection{Frequency-Domain Signal Processing}
\label{subsec:freq_arch}
\label{subsec:time_domain}

The radio interface exchanges complex samples with signed 16-bit real and imaginary components (\texttt{c16}). The emulator applies frequency-domain filtering using overlap-save with transform size $C$ and a configured maximum impulse-response length $L_{\max}$. The current implementation uses $C=4096$. Each transform contains $L_{\max}-1$ preceding input samples and $P=C-L_{\max}+1$ new samples. Let $N$ denote the number of samples per antenna in a processing interval; $N=N_\mathrm{slot}$ when the interval is one radio slot. The transform batch contains $M=\lceil N/P\rceil$ blocks. A GPU kernel converts the input to complex single precision and zero-pads any unused positions in the final block. After channel multiplication and inverse transformation, the output kernel discards the first $L_{\max}-1$ samples of each block and retains exactly $N$ output samples. It also applies the $1/C$ inverse-transform normalization and saturation to the \texttt{c16} range.

Initialization requires $L_{\max}\leq C/2$, with a default of $L_{\max}=512$. The transform count $M$ therefore depends on the interval length and configured filter support. The processing interval has duration $\tau_\mathrm{int}=N/f_s$, where $f_s$ is the sampling rate in samples per second. The emulator processes the complete time-domain waveform, including cyclic prefixes, without decoding OFDM symbols. Its transform bins are therefore processing bins, not necessarily the radio's OFDM subcarriers.

For block $m\in\{0,\ldots,M-1\}$ and FFT bin $q\in\{0,\ldots,C-1\}$, let $X_m[q]$ and $Y_m[q]$ denote the complex input and filtered output spectra, respectively, and let $H[q]$ denote the scalar channel frequency response. The channel operation is
\begin{equation}
    Y_m[q]=H[q]X_m[q], \qquad 0\leq q<C.
    \label{eq:block_channel}
\end{equation}
With a fixed transform size, the FFT work per interval scales as $\mathcal{O}(MC\log C)$ and the scalar multiplication as $\mathcal{O}(MC)$ per link. Direct filtering with an $L$-tap impulse response instead requires $\mathcal{O}(NL)$ operations per link. Section~\ref{subsec:eval_pipeline} compares processing latency with a matched CPU baseline and quantifies the benefit of batched GPU dispatch.

For a fixed channel with at most $L_{\max}$ causal taps, overlap-save produces the same output as direct time-domain convolution, up to floating-point and quantization errors. The emulator maintains input history separately for each input stream across processing intervals. Forward timestamp gaps are represented by zero input; backward or repeated timestamps reset the history. When transmission on a link stops, zero input is processed with the retained input history to emit the remaining channel tail. The response-support check and coefficient-update policy are described in Section~\ref{subsec:channel_model}; Section~\ref{subsec:eval_fidelity} evaluates signal fidelity for the tested configurations.

\subsection{Multi-UE and Multi-Antenna Processing}
\label{subsec:multi_ue}

On the downlink, all UEs receive a channel-filtered version of the gNB waveform. The emulator computes the forward transform once per transmit antenna and retains the result in an immutable device buffer. Each UE's channel application reads this common spectrum and writes to a separate output region. We index the $K$ UEs by $k\in\{1,\ldots,K\}$. Bold symbols denote vectors or matrices. For UE~$k$,
\begin{equation}
    \mathbf{Y}^{\mathrm{DL}}_{k,m}[q]
    =\mathbf{H}_{k}[q]\mathbf{X}^{\mathrm{DL}}_m[q],
    \label{eq:dl_channel}
\end{equation}
where $\mathbf{X}^{\mathrm{DL}}_m[q]\in\mathbb{C}^{n_\mathrm{tx}}$ is the gNB transmit spectrum, $\mathbf{Y}^{\mathrm{DL}}_{k,m}[q]\in\mathbb{C}^{n_\mathrm{rx}}$ is the filtered spectrum at UE~$k$, and $\mathbf{H}_{k}[q]\in\mathbb{C}^{n_\mathrm{rx}\times n_\mathrm{tx}}$ is its channel matrix. Here, $n_\mathrm{tx}$ and $n_\mathrm{rx}$ refer to the downlink antenna roles. SISO reduces these quantities to scalars. The equations describe channel filtering before receiver noise, output gain, and quantization. A GPU kernel applies the channel matrices across UEs and bins, and batched inverse transforms recover one waveform per UE receive antenna. Sharing the input spectra avoids repeating the forward transforms for each UE.

On the uplink, the emulator transforms each participating UE's transmitted waveform and sums the channel-weighted contributions in the frequency domain. For each coefficient version, the MIMO implementation represents uplink reciprocity through the transpose of the downlink matrix:
\begin{equation}
    \mathbf{Y}^{\mathrm{UL}}_m[q]
    =\sum_{k=1}^{K}\mathbf{H}_{k}^{\mathsf{T}}[q]
      \mathbf{X}^{\mathrm{UL}}_{k,m}[q].
    \label{eq:ul_channel}
\end{equation}
Here, $\mathbf{X}^{\mathrm{UL}}_{k,m}[q]\in\mathbb{C}^{n_\mathrm{rx}}$ is the UE transmit spectrum, $\mathbf{Y}^{\mathrm{UL}}_m[q]\in\mathbb{C}^{n_\mathrm{tx}}$ is the combined spectrum at the gNB, and $(\cdot)^{\mathsf{T}}$ denotes transpose. Antenna roles reverse on the uplink. The sum is inverse-transformed once per gNB receive antenna, producing a composite waveform of simultaneous UE transmissions. The current MIMO interface requires equal configured transmit and receive antenna counts. Each equation uses the coefficient version selected for that direction and processing interval; the version index is omitted for brevity. The two directions select versions independently, as described in Section~\ref{subsec:channel_model}.

Receiver noise and output gain are configured separately from the propagation coefficients. Noise power follows thermal noise and a configured receiver noise figure, unless explicitly specified. Physical-network agreement requires separate calibration. The downlink output kernel normalizes the inverse transform, adds \gls{awgn}, applies the per-UE gain in floating point, and quantizes the result. Uplink noise is added after combining UE contributions, representing noise at the gNB receiver.

\subsection{Timestamped Shared-Memory Transport}
\label{subsec:shm_transport}

\begin{figure*}[t]
    \centering
    % Editable, two-column IEEE figure. Single-gNB view; k = 1,...,K.
% All solid arrows carry time-domain c16 IQ. Spectral operations stay on GPU.
\begingroup
\definecolor{iqDL}{HTML}{215A85}
\definecolor{iqUL}{HTML}{A45421}
\definecolor{iqInk}{HTML}{303840}
\definecolor{iqRule}{HTML}{89939C}
\begin{tikzpicture}[
  x=1cm,y=1cm,font=\sffamily\footnotesize,text=iqInk,
  box/.style={draw=iqRule,line width=.6pt,rounded corners=2pt,
              align=center,inner sep=4pt},
  ring/.style={box,minimum width=2.65cm,minimum height=1.02cm},
  dl/.style={-{Latex[length=1.8mm]},draw=iqDL,line width=.85pt},
  ul/.style={-{Latex[length=1.8mm]},draw=iqUL,line width=.85pt},
  control/.style={-{Latex[length=1.8mm]},draw=iqInk,dashed,line width=.7pt}
]
% Fixed bounding box keeps the figure within IEEEtran's two-column width.
\path[use as bounding box] (0,.35) rectangle (17.1,5.1);

% GPU enclosure receives the asynchronous coefficient-update arrow.
\draw[draw=iqRule,rounded corners=3pt,line width=.6pt]
  (5.65,.38) rectangle (11.05,4.05);

% Column headings distinguish processes, transport, and computation.
\node[font=\footnotesize] at (.8,3.7) {Radio stack};
\node[font=\footnotesize] at (3.45,3.7) {Common IQ rings};
\node[font=\footnotesize] at (8.35,3.7) {GPU channel emulator};
\node[font=\footnotesize] at (13.25,3.7) {Per-UE IQ rings};
\node[font=\footnotesize] at (16.15,3.7) {Radio stacks};

% Coefficients are delivered independently of the sample path.
\node[box,fill=black!3,minimum width=6.0cm,minimum height=.65cm]
  (updates) at (8.35,4.7)
  {Channel producer $\rightarrow$ Redis subscriber $\rightarrow$ Versioned CFRs};
\draw[control] (updates.south) -- (8.35,4.05);

% Process blocks span both directional lanes.
\node[box,fill=black!3,minimum width=1.45cm,minimum height=2.65cm]
  (gnb) at (.8,1.9) {gNB\\process};
\node[box,fill=black!3,minimum width=1.7cm,minimum height=2.65cm]
  (ues) at (16.15,1.9) {UE $k$\\process\\[3pt]$k=1,\ldots,K$};

% Downlink lane: common waveform, per-link filtering, individual outputs.
\node[ring,fill=iqDL!6] (dlcommon) at (3.45,2.7)
  {Downlink\\gNB transmit samples};
\node[box,draw=iqDL,fill=iqDL!5,minimum width=5.05cm,
      minimum height=1.15cm] (dlgpu) at (8.35,2.7)
  {Downlink: shared FFT, per-UE filtering\\[2pt]
   Apply channel $\mathbf{H}_k[q]$ $\rightarrow$ IFFT $\rightarrow$ IQ};
\node[ring,fill=iqDL!6] (dlue) at (13.25,2.7)
  {Downlink $k$\\UE receive samples};
\draw[dl] (gnb.east |- dlcommon) -- node[above,font=\scriptsize] {write} (dlcommon.west);
\draw[dl] (dlcommon.east) -- (dlgpu.west);
\draw[dl] (dlgpu.east) -- (dlue.west);
\draw[dl] (dlue.east) -- node[above,font=\scriptsize] {read} (ues.west |- dlue);

% Uplink lane: raw UE samples enter the GPU; channel weighting occurs here.
\node[ring,fill=iqUL!6] (ulcommon) at (3.45,1.1)
  {Uplink\\gNB receive samples};
\node[box,draw=iqUL,fill=iqUL!5,minimum width=5.05cm,
      minimum height=1.15cm] (ulgpu) at (8.35,1.1)
  {Uplink: per-UE FFT, channel and sum\\[2pt]
   Combine UE signals $\rightarrow$ IFFT $\rightarrow$ IQ};
\node[ring,fill=iqUL!6] (ulue) at (13.25,1.1)
  {Uplink $k$\\UE transmit samples};
\draw[ul] (ues.west |- ulue) -- node[above,font=\scriptsize] {write} (ulue.east);
\draw[ul] (ulue.west) -- (ulgpu.east);
\draw[ul] (ulgpu.west) -- (ulcommon.east);
\draw[ul] (ulcommon.west) -- node[above,font=\scriptsize] {read} (gnb.east |- ulcommon);

\end{tikzpicture}
\endgroup
    \caption{IQ transport and GPU processing for one gNB and $K$ UEs. Downlink processing applies per-UE channels to a common gNB waveform; uplink processing combines channel-filtered UE waveforms. Shared-memory rings carry time-domain IQ samples, while CFR updates arrive asynchronously.}
    \label{fig:shm}
\end{figure*}
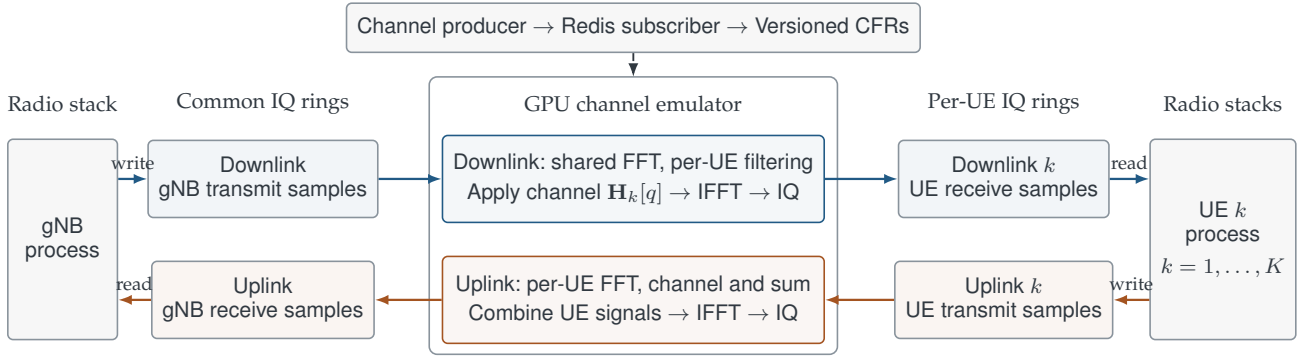

Stack-specific \texttt{gpuemu} radio drivers connect the radio stacks to a common shared-memory interface. The OAI driver uses the \texttt{openair0} read and write callbacks; the OCUDU plugin implements its radio-session interface (Section~\ref{subsec:ocudu}). The gNB writes its downlink samples to a common \gls{shm} segment, and the emulator writes each filtered waveform to a per-UE downlink segment. In the reverse direction, each UE writes to its own uplink segment, and the emulator publishes the composite waveform to the gNB's uplink segment. This requires $2(K+1)$ IQ segments for one gNB and $K$ UEs (Fig.~\ref{fig:shm}). Antenna streams occupy separate planes within each segment.

Each antenna plane stores $B=8{,}601{,}600$ complex samples, providing $B/f_s$ seconds of buffering at sampling rate $f_s$. Sample $n$ is stored at ring-buffer offset $n\bmod B$. The producer writes all antenna samples before advancing the publication timestamp. Readers validate requested intervals against this timestamp and the ring capacity. An atomic sequence counter and Linux futex wakeups provide interprocess notification.

The emulator registers shared-memory sample planes with CUDA for asynchronous host--device transfers, removing an intermediate host copy. Valid, contiguous ranges use direct transfers; wraparound, incomplete input, or unavailable registration selects a pinned staging buffer. Output is published only after the device-to-host transfer completes, preventing readers from accessing unfinished GPU output.

\subsection{Slot Timing and Uplink Collection}
\label{subsec:timing}

Separate host threads process downlink and uplink samples. The downlink thread waits for samples produced by the gNB. The uplink thread is paced by the host's monotonic clock, with successive deadlines separated by the processing-interval duration $\tau_\mathrm{int}$. It also checks the downlink processing timestamp and waits if its own timestamp is more than one interval ahead. By default, this wait continues until downlink processing catches up, so a downlink stall can also stop uplink progress. An optional timeout allows uplink processing to continue despite that stall. These rules set the intended processing pace; meeting the deadlines still depends on host scheduling and processing time.

For each uplink interval, one worker per UE waits for that UE's samples. The workers run concurrently and account for the offset between UE transmit timestamps and the gNB receive window, including timing advance. These sample-collection waits are separate from the downlink-progress wait described above. If a collection wait expires, the collector can use the samples already available at the start of the interval and fill the missing portion with zeros. A link with no available samples contributes zero input, although its retained filter history may still produce a channel tail. Concurrent collection avoids adding a separate wait for every UE; completion depends on the last worker to finish and the configured timeouts.

During GPU initialization, a temporary thread advances the gNB-facing uplink publication timestamp, which identifies how far samples have been made available to the gNB. The emulator then selects a starting timestamp from the gNB downlink position and advances lagging ring timestamps to that reference. This establishes the initial sample-counter positions but can leave an offset between those counters and the actual sample content.

If downlink processing falls behind and resumes at a newer timestamp, the emulator first clears the skipped ranges in the UE output rings, then advances their publication timestamps. Readers therefore receive zeros for those ranges rather than stale samples from an earlier traversal of the ring.

\subsection{OCUDU Integration}
\label{subsec:ocudu}

\mirage also supports an OCUDU gNB communicating with OAI software UEs. A radio driver connects OCUDU to the same timestamped IQ interface used by OAI, while preserving the timing required by its physical layer. The integration reuses the channel emulator without modification and requires only build and configuration adaptations for the OAI UEs. This separation allows different gNB stacks to use the same channel-processing framework.

\subsection{GPU Execution and Resource Management}
\label{subsec:memory_arch}
\label{subsec:deployment}

Per-antenna FFT plans, device workspaces, and pinned staging buffers are allocated during initialization. The multi-UE inverse-transform plan is created on first use and recreated when the UE count changes. Preallocation avoids repeated workspace allocation during steady-state processing, while ping-pong buffers separate successive submissions.

Downlink processing batches channel application and inverse transforms across UEs. Uplink processing orders input transforms, matrix mixing, and output transforms through CUDA streams and synchronization. The host waits for output completion before publishing samples. CUDA's blocking synchronization policy allows waiting host threads to yield the CPU.

\subsection{Asynchronous Channel Updates}
\label{subsec:channel_model}

The channel producer sends per-link CFRs through Redis independently of IQ processing. Each update identifies the link, version, antenna dimensions, and frequency grid and contains complex single-precision coefficients. The producer and emulator must use matching carrier frequencies, sample rates, and antenna layouts. The subscriber validates each update and converts its frequency-bin ordering when needed.

Before uploading a response, the emulator rejects nonfinite coefficients and checks the impulse-response energy beyond $L_{\max}$ samples. If that energy fraction exceeds the configured tolerance (default $10^{-3}$), the update is rejected. Otherwise, tails above numerical precision are removed and the CFR is reconstructed. Disabling this check removes the linear-convolution guarantee for longer responses. These checks run outside the IQ-processing loop.

Each link uses four GPU coefficient buffers. A new version becomes available only after its upload completes. IQ processing continues with the current coefficients during this upload; it does not wait for the next channel response. At the start of each interval, each direction selects the latest available version for each link and uses it throughout that interval. A buffer is reused only after both directions finish using it, preventing partially updated coefficients from entering the filtering operation.

Downlink, uplink, and different links select versions independently, so updates are not simultaneous across the network. Intermediate versions may be replaced before use, and coefficients are not interpolated. This mechanism ensures complete updates; their freshness depends on the producer and delivery delay.

\section{Evaluation}
\label{sec:eval}
We evaluate \mirage on an H100 platform through component tests of processing time, waveform accuracy, and channel-update consistency, followed by full-stack delivery and concurrent-instance experiments. We check whether processing finishes within each slot, whether traffic is delivered, and whether the radio clock keeps pace with wall time.
% Common, print-readable styles for Section V only.
\definecolor{evCPU}{HTML}{666666}
\definecolor{evFFT}{HTML}{0072B2}
\definecolor{evTD}{HTML}{D55E00}
\definecolor{evMIMOtwo}{HTML}{009E73}
\definecolor{evMIMOfour}{HTML}{CC79A7}
\definecolor{evGold}{HTML}{E69F00}
\pgfplotsset{
 eval axis/.style={
   width=0.94\linewidth, height=4.6cm,
   font=\footnotesize, tick label style={font=\scriptsize},
   label style={font=\footnotesize},
   grid=major, major grid style={solid,black!12,line width=0.2pt},
   minor tick num=0, tick align=outside,
   axis line style={black!70,line width=0.4pt},
   tick style={black!70,line width=0.4pt},
   line width=0.8pt, mark size=1.8pt,
   legend cell align=left,
   legend style={font=\scriptsize,draw=none,fill=white,inner sep=2pt,
      at={(0.5,1.04)},anchor=south,legend columns=2,column sep=5pt},
   error bars/error bar style={line width=0.5pt},
   unbounded coords=jump,
 },
 eval cpu/.style={color=evCPU,mark=o,solid,mark options={solid,draw=evCPU,fill=white}},
 eval td/.style={color=evTD,mark=triangle*,dashed,mark options={solid,draw=evTD,fill=evTD}},
 eval fft/.style={color=evFFT,mark=square*,solid,mark options={solid,draw=evFFT,fill=evFFT}},
 eval mimo2/.style={color=evMIMOtwo,mark=diamond*,dashdotted,mark options={solid,draw=evMIMOtwo,fill=evMIMOtwo}},
 eval mimo4/.style={color=evMIMOfour,mark=pentagon*,dotted,mark options={solid,draw=evMIMOfour,fill=evMIMOfour}},
}

\subsection{Platform and Measurement Methodology}
\label{subsec:eval_h100}

Experiments use one NVIDIA H100 SXM GPU with 80~GB of memory on a dual-socket Intel Xeon Platinum 8468 host and the OAI 2026 week~38 release. Component tests use synthetic radio peers; full-stack tests execute OAI gNB and UE stacks unless explicitly identified as OCUDU SISO tests in Section~\ref{subsec:eval_ocudu}. In single-gNB tests, $K$ denotes the number of gNB--UE links.

We focus on FR1, using 30~kHz SCS for full-stack tests and 15 and 30~kHz for component comparisons, consistent with OAI support~\cite{oai_ran_features}. We omit 60~kHz because our timing test uses equal-sized blocks, whereas OAI alternates slot sample counts at this SCS. The 120~kHz setting is outside our FR1 scope.

Each component timing configuration uses five independent process starts, a 30~s warm-up, and a 60~s measurement window, with channel updates at 10~Hz. Tests run in isolation with real-time scheduling for processing threads. GPU assignment, thread priorities, and CPU affinity are verified before full-stack traffic and after cell restarts.

The waveform tests include two operations that prevent old or missing samples from reaching the receiver: clearing skipped intervals in the sample buffer and returning zeros when a driver read fails. Processing time is measured without these two operations. Their added processing time has not been measured.

\subsection{Processing Time and Scaling}
\begin{figure}[t]
\centering
% Shared legend for the paired panels.
\begin{tikzpicture}
\begin{axis}[hide axis,scale only axis,width=1mm,height=1mm,
xmin=0,xmax=1,ymin=0,ymax=1,
legend columns=2,legend cell align=left,
legend style={font=\scriptsize,draw=none,inner sep=1pt,column sep=3pt,
at={(0,0)},anchor=center}]
\addlegendimage{eval cpu}
\addlegendentry{61.44 Msps, SISO}
\addlegendimage{eval fft}
\addlegendentry{122.88 Msps, SISO}
\addlegendimage{eval mimo2}
\addlegendentry{61.44 Msps, $2\times2$}
\addlegendimage{eval td}
\addlegendentry{122.88 Msps, $2\times2$}
\addlegendimage{eval mimo4}
\addlegendentry{61.44 Msps, $4\times4$}
\end{axis}
\end{tikzpicture}\par\smallskip
\subfloat[Downlink]{\begin{minipage}[b]{\linewidth}\centering
% Generated by scripts/gen_h100_snapshot.py.
\begin{tikzpicture}
\begin{axis}[eval axis,width=0.96\linewidth,height=4.5cm,
xmode=log,log basis x=2,log ticks with fixed point,xtick={1,2,4,8,16},xmin=0.9,xmax=48,
xlabel={Number of links, $K$},ylabel={Per-run p99 ($\mu$s)},ymin=0,ymax=1150,ytick={0,250,500,750,1000},
legend style={legend columns=2,font=\scriptsize,column sep=3pt,row sep=1pt,inner sep=1pt,at={(0.5,1.03)},anchor=south}]
\addplot+[eval cpu,error bars/y dir=both,error bars/y explicit] table[col sep=comma,x=K,y=p99,y error minus=lo,y error plus=hi] {data/tikz/h100_20261008/dl_61.44_1.csv};

\addplot+[eval fft,error bars/y dir=both,error bars/y explicit] table[col sep=comma,x=K,y=p99,y error minus=lo,y error plus=hi] {data/tikz/h100_20261008/dl_122.88_1.csv};

\addplot+[eval mimo2,error bars/y dir=both,error bars/y explicit] table[col sep=comma,x=K,y=p99,y error minus=lo,y error plus=hi] {data/tikz/h100_20261008/dl_61.44_2.csv};

\addplot+[eval td,error bars/y dir=both,error bars/y explicit] table[col sep=comma,x=K,y=p99,y error minus=lo,y error plus=hi] {data/tikz/h100_20261008/dl_122.88_2.csv};

\addplot+[eval mimo4,error bars/y dir=both,error bars/y explicit] table[col sep=comma,x=K,y=p99,y error minus=lo,y error plus=hi] {data/tikz/h100_20261008/dl_61.44_4.csv};

% Reference budgets; measurements use 30 kHz SCS.
\addplot[black!45,densely dotted,line width=0.5pt,mark=none,forget plot] coordinates {(0.9,1000) (48,1000)};
\addplot[black,dashed,line width=0.8pt,mark=none,forget plot] coordinates {(0.9,500) (48,500)};
\node[anchor=north east,font=\scriptsize,text=black!65,fill=white,inner sep=1pt] at (axis cs:46,1000) {15 kHz: 1000 $\mu$s};
\node[anchor=south east,font=\scriptsize,fill=white,inner sep=1pt] at (axis cs:46,500) {30 kHz: 500 $\mu$s};
\end{axis}
\end{tikzpicture}

\end{minipage}}\par\smallskip
\subfloat[Uplink]{\begin{minipage}[b]{\linewidth}\centering
% Generated by scripts/gen_h100_snapshot.py.
\begin{tikzpicture}
\begin{axis}[eval axis,width=0.96\linewidth,height=4.5cm,
xmode=log,log basis x=2,log ticks with fixed point,xtick={1,2,4,8,16},xmin=0.9,xmax=48,
xlabel={Number of links, $K$},ylabel={Per-run p99 ($\mu$s)},ymin=0,ymax=1150,ytick={0,250,500,750,1000},
legend style={legend columns=2,font=\scriptsize,column sep=3pt,row sep=1pt,inner sep=1pt,at={(0.5,1.03)},anchor=south}]
\addplot+[eval cpu,error bars/y dir=both,error bars/y explicit] table[col sep=comma,x=K,y=p99,y error minus=lo,y error plus=hi] {data/tikz/h100_20261008/ul_61.44_1.csv};

\addplot+[eval fft,error bars/y dir=both,error bars/y explicit] table[col sep=comma,x=K,y=p99,y error minus=lo,y error plus=hi] {data/tikz/h100_20261008/ul_122.88_1.csv};

\addplot+[eval mimo2,error bars/y dir=both,error bars/y explicit] table[col sep=comma,x=K,y=p99,y error minus=lo,y error plus=hi] {data/tikz/h100_20261008/ul_61.44_2.csv};

\addplot+[eval td,error bars/y dir=both,error bars/y explicit] table[col sep=comma,x=K,y=p99,y error minus=lo,y error plus=hi] {data/tikz/h100_20261008/ul_122.88_2.csv};

\addplot+[eval mimo4,error bars/y dir=both,error bars/y explicit] table[col sep=comma,x=K,y=p99,y error minus=lo,y error plus=hi] {data/tikz/h100_20261008/ul_61.44_4.csv};

% Reference budgets; measurements use 30 kHz SCS.
\addplot[black!45,densely dotted,line width=0.5pt,mark=none,forget plot] coordinates {(0.9,1000) (48,1000)};
\addplot[black,dashed,line width=0.8pt,mark=none,forget plot] coordinates {(0.9,500) (48,500)};
\node[anchor=north east,font=\scriptsize,text=black!65,fill=white,inner sep=1pt] at (axis cs:46,1000) {15 kHz: 1000 $\mu$s};
\node[anchor=south east,font=\scriptsize,fill=white,inner sep=1pt] at (axis cs:46,500) {30 kHz: 500 $\mu$s};
\end{axis}
\end{tikzpicture}

\end{minipage}}
\caption{Processing time versus link count at 30~kHz SCS. Error bars show 95\% confidence intervals. Horizontal lines mark the labeled slot budgets.}
\label{fig:h100_timing}
\end{figure}
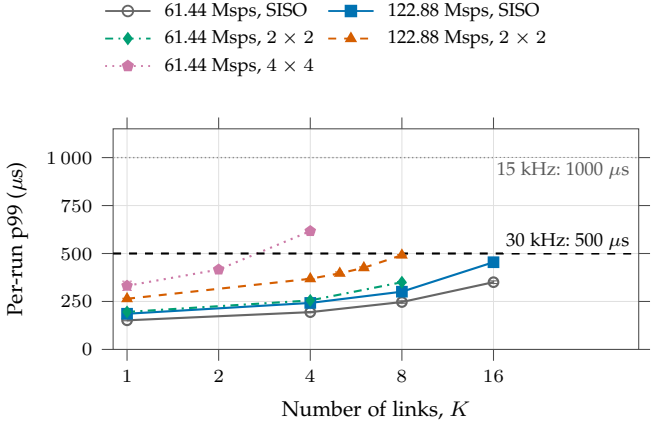
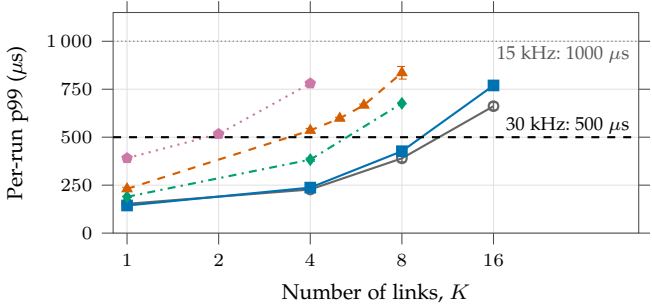

\begin{figure}[t]\centering
\begin{tikzpicture}\begin{axis}[eval axis,width=0.96\linewidth,height=3.5cm,ymin=0,ymax=1.05,xtick={1,2,3,4},xticklabels={15/1,15/4,30/1,30/4},xlabel={SCS (kHz) / links},ylabel={p99 processing / slot duration}]
\addplot+[eval fft,error bars/.cd,y dir=both,y explicit] table[col sep=comma,x=x,y=dl,y error minus expr=\thisrow{dl}-\thisrow{dllo},y error plus expr=\thisrow{dlhi}-\thisrow{dl}] {data/tikz/h100_final/numerology.csv};\addlegendentry{DL}
\addplot+[eval mimo2,error bars/.cd,y dir=both,y explicit] table[col sep=comma,x=x,y=ul,y error minus expr=\thisrow{ul}-\thisrow{ullo},y error plus expr=\thisrow{ulhi}-\thisrow{ul}] {data/tikz/h100_final/numerology.csv};\addlegendentry{UL}
\end{axis}\end{tikzpicture}
\caption{Processing time relative to the slot budget at 15 and 30~kHz SCS. Points show mean per-run p99 ratios with 95\% confidence intervals.}
\label{fig:h100_numerology}
\end{figure}
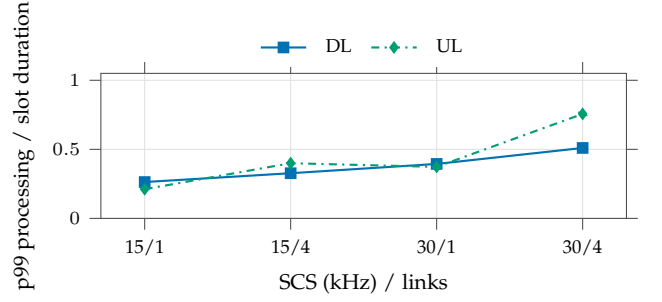

Figure~\ref{fig:h100_timing} reports the mean of the five per-run 99th-percentile processing times, with 95\% Student-$t$ confidence intervals across starts. Processing time covers input collection, GPU work, and output publication within each interval; it does not measure the full delay from input availability to output publication. Buffering and startup alignment can add delay. Downlink statistics include all measured slots, whereas uplink statistics include slots with a UE write. The time available to process $N$ samples at sampling rate $f_s$ is $\tau_\mathrm{int}=N/f_s$. When one interval covers one NR slot, its nominal duration is $1/2^\mu$~ms, where the subcarrier spacing is $15\times2^\mu$~kHz~\cite{3gpp38211}. Figure~\ref{fig:h100_timing} marks the nominal budgets for 15 and 30~kHz SCS. The plotted measurements use 30~kHz SCS, so their processing budget is $500~\mu$s. For comparisons across SCS values, Fig.~\ref{fig:h100_numerology} reports p99 processing time divided by the interval duration; a value of one means the full budget is used. A configuration meets the processing criterion only when both directions have p99 at most $\tau_\mathrm{int}$ and fewer than 1\% of intervals exceed that budget in every start.

Uplink processing limits the tested operating range (Table~\ref{tab:h100_envelope}). At $61.44$~Msps with four $2\times2$ links, mean per-run p99 is approximately $255~\mu$s downlink and $383~\mu$s uplink. At $122.88$~Msps with the same link and antenna counts, it increases to $368~\mu$s and $535~\mu$s, respectively; the mean fraction of active uplink intervals above budget is 19.9\%. Thus, a downlink-only timing assessment would overstate the bidirectional operating range. The eight-link, $122.88$~Msps $2\times2$ case reaches $836~\mu$s uplink p99.

\begin{table}[t]
\centering
\caption{Highest passing link count among tested configurations.}
\label{tab:h100_envelope}
\small
\begin{tabular}{@{}lcc@{}}\toprule
Rate (Msps) & Antennas & Highest passing $K$\\\midrule
61.44 & SISO & 8\\
61.44 & $2\times2$ & 4\\
61.44 & $4\times4$ & 1\\
122.88 & SISO & 8\\
122.88 & $2\times2$ & 1\\\bottomrule
\end{tabular}
\end{table}

All tested configurations meet the harness delivery criterion, including those that exceed the processing budget. The longer receive wait and TDD traffic pattern allow delivery despite active uplink intervals exceeding one slot.

At 61.44~Msps with $2\times2$ links, Fig.~\ref{fig:h100_numerology} compares 1~ms and 0.5~ms intervals over five starts, with approximately matched TDD time allocation. The shorter interval reduces the batch size but increases the fraction of the budget used by uplink processing at four links. This comparison uses synthetic peers rather than full-stack operation.

\subsection{Matched Computational Baseline}
\label{subsec:eval_ablation}
\label{subsec:eval_pipeline}
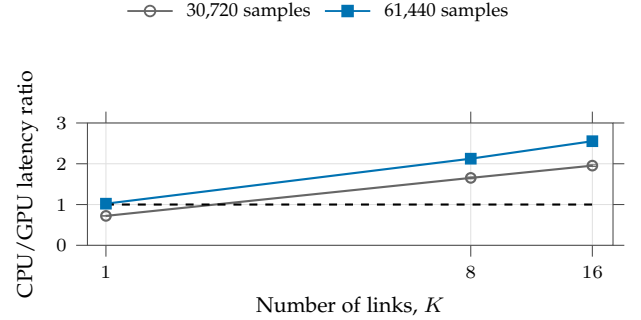
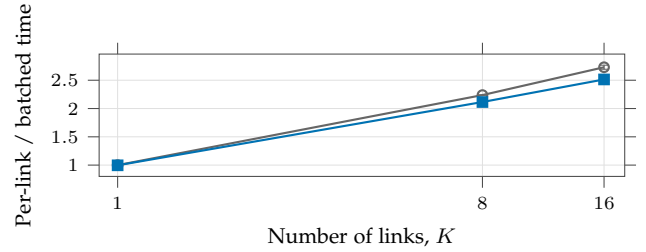
\begin{figure}[t]\centering
% Shared legend for the paired panels.
\begin{tikzpicture}
\begin{axis}[hide axis,scale only axis,width=1mm,height=1mm,
xmin=0,xmax=1,ymin=0,ymax=1,
legend columns=2,legend cell align=left,
legend style={font=\scriptsize,draw=none,inner sep=1pt,column sep=3pt,
at={(0,0)},anchor=center}]
\addlegendimage{eval cpu}
\addlegendentry{30,720 samples}
\addlegendimage{eval fft}
\addlegendentry{61,440 samples}
\end{axis}
\end{tikzpicture}\par\smallskip
\subfloat[GPU versus six-core FFTW baseline]{\begin{minipage}[b]{\linewidth}\centering
% Generated by scripts/gen_h100_snapshot.py.
\begin{tikzpicture}
\begin{axis}[eval axis,width=0.96\linewidth,height=3.2cm,
xmode=log,log basis x=2,log ticks with fixed point,xtick={1,8,16},xmin=0.9,xmax=18,
xlabel={Number of links, $K$},ylabel={CPU/GPU latency ratio},ymin=0,ymax=3]
\addplot+[eval cpu,error bars/y dir=both,error bars/y explicit] table[col sep=comma,x=K,y=ratio,y error minus=lo,y error plus=hi] {data/tikz/h100_20261008/speedup_30720.csv};

\addplot+[eval fft,error bars/y dir=both,error bars/y explicit] table[col sep=comma,x=K,y=ratio,y error minus=lo,y error plus=hi] {data/tikz/h100_20261008/speedup_61440.csv};

\addplot[black,dashed,mark=none,forget plot] coordinates {(1,1) (16,1)};
\end{axis}
\end{tikzpicture}
\end{minipage}}\par\smallskip
\subfloat[Batched versus per-link GPU dispatch]{\begin{minipage}[b]{\linewidth}\centering
\begin{tikzpicture}\begin{axis}[eval axis,width=0.96\linewidth,height=3.2cm,xmode=log,log basis x=2,log ticks with fixed point,xmin=0.9,xmax=18,xtick={1,8,16},xlabel={Number of links, $K$},ylabel={Per-link / batched time},ymin=0.8,legend columns=1]
\addplot+[eval cpu,error bars/.cd,y dir=both,y explicit] table[col sep=comma,x=K,y=ratio,y error minus expr=\thisrow{ratio}-\thisrow{lo},y error plus expr=\thisrow{hi}-\thisrow{ratio}] {data/tikz/h100_final/batching_30720.csv};
\addplot+[eval fft,error bars/.cd,y dir=both,y explicit] table[col sep=comma,x=K,y=ratio,y error minus expr=\thisrow{ratio}-\thisrow{lo},y error plus expr=\thisrow{hi}-\thisrow{ratio}] {data/tikz/h100_final/batching_61440.csv};
\end{axis}\end{tikzpicture}
\end{minipage}}
\caption{Matched computational comparisons for 64-tap SISO filtering. Values above one favor batched GPU processing. Points show geometric means of five paired median-latency ratios; whiskers show 95\% confidence intervals.}
\label{fig:h100_speedup}
\end{figure}

Figure~\ref{fig:h100_speedup} compares GPU filtering with a matched six-core FFTW CPU baseline for 64-tap SISO channels. For each of five paired runs, we divide the median CPU time by the median GPU time. We report the geometric mean of these ratios and its 95\% confidence interval. At one link, the CPU is faster for 30,720 samples and approximately equal for 61,440 samples. At 16 links, GPU acceleration reaches approximately $1.95\times$ and $2.55\times$, respectively. A separate eight-link $2\times2$ case at 61,440 samples reaches $2.93\times$ (95\% interval: $2.90$--$2.95$). GPU acceleration therefore increases with workload size in these tests.

Batching reduces dispatch overhead as the number of links increases. At 16 SISO links, the per-link-to-batched time ratio is $2.73\times$ for 30,720 samples and $2.51\times$ for 61,440 samples. At the latter size, the CPU-to-GPU ratio varies only from $2.55\times$ to $2.57\times$ across the tested 16--128-tap channels. This limited variation applies to the tested filter lengths and fixed workload.

\subsection{Waveform Accuracy and Channel Updates}
\label{subsec:eval_fidelity}
\label{subsec:eval_updates}
We check waveform accuracy by comparing the emulator output with the expected samples for known channel responses. Tests cover SISO and $2\times2$ links at 61.44 and 122.88~Msps, in both directions, with one link and five random seeds per configuration. Across 160 tests, identity and delayed-impulse channels match exactly. Dense long-support and near-saturation channels differ by at most one integer quantization step per real or imaginary component.

We also check that each processing interval uses a complete channel response, without mixing old and new coefficients. SISO and $2\times2$ tests use 10 and 100 updates/s, five runs per configuration, and a 5~ms uplink receive wait. No mixed updates are detected across 16.7 million analyzed slots, and every received update is applied.

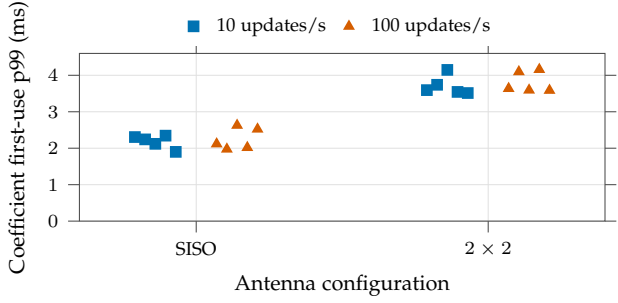
\begin{figure}[t]
\centering
% Generated by scripts/gen_h100_update_delay.py from F4B_REPORT.json.
\begin{tikzpicture}
\begin{axis}[eval axis,width=0.96\linewidth,height=3.8cm,
xmin=0.6,xmax=2.4,xtick={1,2},xticklabels={SISO,$2\times2$},
xlabel={Antenna configuration},ylabel={Coefficient first-use p99 (ms)},
ymin=0,ymax=4.6,ytick={0,1,2,3,4},legend columns=2]
\addplot+[eval fft,only marks,x filter/.code={\ifnum\thisrow{rate_hz}=10\relax\else\def\pgfmathresult{nan}\fi}] table[col sep=comma,x=x,y=p99_ms] {data/tikz/h100_final/update_delay_starts.csv};
\addlegendentry{10 updates/s}
\addplot+[eval td,only marks,x filter/.code={\ifnum\thisrow{rate_hz}=100\relax\else\def\pgfmathresult{nan}\fi}] table[col sep=comma,x=x,y=p99_ms] {data/tikz/h100_final/update_delay_starts.csv};
\addlegendentry{100 updates/s}
\end{axis}
\end{tikzpicture}
\caption{Delay before new channel coefficients are first used. Points show each run's largest stream-level p99.}
\label{fig:h100_update_delay}
\end{figure}

Figure~\ref{fig:h100_update_delay} reports the first-use delay for supplied channel coefficients. The tests use synthetic coefficient updates and do not measure channel-model computation, such as ray tracing. For each run, we plot the largest p99 delay among the gNB uplink and the two UE downlink streams. Values range from 1.90 to 2.63~ms for SISO and from 3.51 to 4.16~ms for $2\times2$. The ranges at 10 and 100 updates/s overlap in both antenna configurations. IQ processing continues with the current coefficients while the next response is received and uploaded. The emulator selects a complete response at the start of a processing interval, as described in Section~\ref{subsec:channel_model}. Thus, the update delay does not represent a pause in IQ processing. These tests do not measure the processing overhead of asynchronous updates.

\subsection{Full-Stack Traffic Delivery}
\label{subsec:eval_fullstack}
We emphasize $2\times2$ operation because it matches the antenna configuration used in our X5G laboratory deployment (Section~\ref{subsec:dt_setup}). Full-stack tests use a 100~MHz carrier, 273 resource blocks, 30~kHz subcarrier spacing, and $2\times2$ antennas. Each configuration uses five independent cell starts, a 60~s warm-up, and sequential 120~s downlink, uplink, and bidirectional tests. Per-slot instrumentation is disabled, and the gNB uplink receive wait is 5~ms. Receiver-side iperf3 UDP reports provide goodput and packet loss. Results exclude repeated tests after a restart.

\begin{table}[t]\centering\small
\caption{Receiver goodput (Mbps): mean and 95\% confidence interval.}
\label{tab:h100_goodput}
\begin{tabular}{@{}rrrc@{}}\toprule
$K$ & Dir. & Offer & Goodput\\\midrule
1 & DL & 20 & $19.87\pm0.06$\\
1 & UL & 32 & $31.91\pm0.23$\\
2 & DL & 80 & $79.85\pm0.36$\\
2 & UL & 64 & $63.98\pm0.01$\\
4 & DL & 160 & $159.89\pm0.05$\\
4 & UL & 64 & $63.98\pm0.01$\\
8 & DL & 320 & $319.47\pm1.01$\\
8 & UL & 32 & $32.00\pm0.00$\\
\bottomrule

\end{tabular}
\end{table}

Table~\ref{tab:h100_goodput} shows that mean goodput closely tracks the offered load for runs with all receiver records available. Each configuration is tested over five independent cell starts; averages and Student-$t$ confidence intervals use the available complete receiver records. At four UEs, all five runs provide receiver records in each direction, delivering 159.89~Mbps downlink and 63.98~Mbps uplink under offers of 160 and 64~Mbps. At eight UEs, mean downlink goodput is 319.47~Mbps under a 320~Mbps offer, based on two runs. These measurements demonstrate traffic delivery at the tested loads; they do not establish maximum throughput.

We also evaluate simultaneous downlink and uplink traffic with four UEs. Across five runs with all receiver records available, mean goodput is 79.71~Mbps downlink and 31.88~Mbps uplink under aggregate offers of 80 and 32~Mbps, respectively. These results complement the separate-direction measurements in Table~\ref{tab:h100_goodput}.

To assess operation across cell starts, we require all flow records, at least 95\% of the offered rate per flow, aggregate packet loss no greater than 1\%, and no detected synchronization or connectivity failure. At four UEs, all five starts meet these criteria in each separate direction, and three meet them under bidirectional traffic. At eight UEs, the corresponding counts are two, four, and two. Synchronization failures also occur when channel filtering is bypassed, so the tests do not identify their cause. Goodput from the available receiver records therefore does not establish reliable operation across all starts.

\subsection{Channel Fidelity from SRS Measurements}
\label{subsec:eval_mimo}
We compare the applied uplink channel with channel estimates obtained from sounding reference signals (SRS) at the OAI gNB. Tests use one UE, 106 resource blocks, 30~kHz SCS, and 61.44~Msps sampling. Identity, delayed single-path, and two-path SISO channels and a $2\times2$ two-path channel each use five independent starts. SRS observations are collected every 80~ms at six uplink noise levels and with noise disabled. SNR is calibrated from occupied and empty SRS comb resource elements. Before comparing the responses, we remove one common complex gain and delay per observation, fitted jointly across all sounded subcarriers and, for $2\times2$, all four channel entries. The comparison therefore evaluates channel shape rather than absolute gain, phase, or propagation delay.

\begin{figure}[t]
\centering
\begin{tikzpicture}
\begin{axis}[eval axis,width=0.97\linewidth,height=3.8cm,
xlabel={Baseband frequency (MHz)},ylabel={$|H|$ (dB)},
xmin=-20,xmax=20,xtick={-20,-10,0,10,20},ymin=-18,ymax=3,ytick={-15,-10,-5,0},
legend columns=2,legend style={column sep=3pt}]
\addplot[color=evFFT,no marks] table[col sep=comma,x=freq,y=applied] {data/tikz/chanval/siso.csv};
\addlegendentry{S-2P applied}
\addplot[color=evFFT,only marks,mark repeat=12,mark=o,mark size=1.3pt,mark options={line width=0.3pt,fill=white}] table[col sep=comma,x=freq,y=estimated] {data/tikz/chanval/siso.csv};
\addlegendentry{S-2P estimate}
\addplot[color=evMIMOtwo,no marks] table[col sep=comma,x=freq,y=applied] {data/tikz/chanval/mimo00.csv};
\addlegendentry{M-2P (0,0) applied}
\addplot[color=evMIMOtwo,only marks,mark repeat=12,mark=o,mark size=1.3pt,mark options={line width=0.3pt,fill=white}] table[col sep=comma,x=freq,y=estimated] {data/tikz/chanval/mimo00.csv};
\addlegendentry{M-2P (0,0) estimate}
\addplot[color=evTD,no marks] table[col sep=comma,x=freq,y=applied] {data/tikz/chanval/mimo10.csv};
\addlegendentry{M-2P (1,0) applied}
\addplot[color=evTD,only marks,mark repeat=12,mark=o,mark size=1.3pt,mark options={line width=0.3pt,fill=white}] table[col sep=comma,x=freq,y=estimated] {data/tikz/chanval/mimo10.csv};
\addlegendentry{M-2P (1,0) estimate}
\end{axis}
\end{tikzpicture}
\caption{Applied (lines) and SRS-estimated (markers) channel magnitude without added noise: two-path SISO (S-2P) and two $2\times2$ entries (M-2P). Common receiver gain is removed.}
\label{fig:srs_channel_fidelity}
\end{figure}
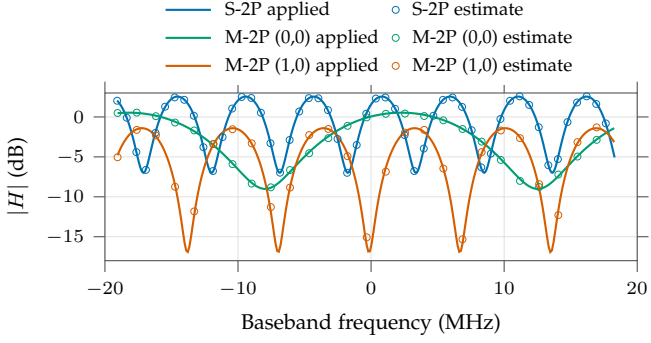

Figure~\ref{fig:srs_channel_fidelity} compares the applied response with a representative SRS observation for each channel. The $2\times2$ entries are indexed by receive antenna and UE port; their references average each subcarrier pair sharing one estimate. The estimated peaks and notches follow the applied responses across the sounded band.

The channel-estimation error follows the expected inverse-SNR trend within 0.13~dB for SISO and within 0.1~dB for $2\times2$ up to 16~dB SNR. Measured noise power is $1.00\pm0.01$ times the configured value, and receiver gain varies by no more than 0.02~dB. With injected noise disabled, the error floors are approximately $-43$~dB for SISO and $-33$~dB for $2\times2$. These residual errors limit the agreement at high SNR; their causes are not fully established.

The seven-sample channel delay is affected by the UE timing-advance loop, so these measurements do not directly validate absolute propagation delay. At the lowest SNR, some SISO runs undergo re-access; the analysis retains observations under the dominant UE identifier, reducing the available samples. Trace collection takes approximately 5~$\mu$s per SRS occasion for SISO and 17~$\mu$s for $2\times2$. Comparison with one trace-disabled start per antenna mode shows no additional attributable deadline misses. The results support agreement between the applied channel shape and the receiver estimate under the tested conditions.

\subsection{Concurrent Full-Stack Instances}
\label{subsec:eval_multiverse}
We test $J=1,2,4$ independent cells on one H100. Each cell uses 40~MHz bandwidth, 30~kHz SCS, $2\times2$ antennas, an identity channel, and two UEs receiving 20~Mbps each. Dedicated physical cores are assigned to the emulator, gNB, and each UE (3/4/2), with four additional cores for the core network. Each emulator uses a dedicated \texttt{1g.20gb} Multi-Instance GPU (MIG) partition with 26 streaming multiprocessors and 20~GB of memory. The partition size is fixed across configurations; unused partitions remain idle.

\begin{figure}[t]\centering
\begin{tikzpicture}
\begin{axis}[eval axis,height=4cm,width=0.97\linewidth,xmin=0.7,xmax=3.4,ymin=0,ymax=1.25,unbounded coords=discard,
xtick={1,2,3},xticklabels={1,2,4},xlabel={Concurrent cells},ylabel={p99 / slot budget},legend columns=2]
\addplot[black!60,dashed,forget plot] coordinates {(0.7,1)(3.4,1)};
\addplot+[eval fft,only marks] table[col sep=comma,x expr=\thisrow{x}-0.07,y=dl] {data/tikz/mpsscale/timing.csv};\addlegendentry{DL}
\addplot+[eval td,only marks] table[col sep=comma,x expr=\thisrow{x}+0.07,y=ul] {data/tikz/mpsscale/timing.csv};\addlegendentry{UL}
\end{axis}\end{tikzpicture}
\caption{Per-cell p99 relative to the slot budget across five completed deployments. All cells use dedicated MIG partitions; the dashed line marks the budget.}
\label{fig:h100_instances}
\end{figure}
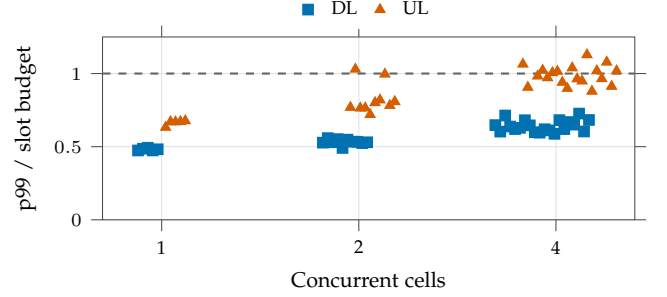

After a 15~s warm-up, traffic runs for 60~s. Figure~\ref{fig:h100_instances} shows five completed deployments per configuration. Starts with lost UEs or failed flows are repeated at most twice; five, seven, and nine starts are needed at $J=1,2,4$, respectively. Each cell must meet the processing and delivery criteria above, advance within 1\% of real time, and accumulate at most 1~ms of backlog. UL timing includes intervals with processed samples; receive timeouts are counted separately.

Single-cell deployments meet all criteria in all five runs, with UL p99 of 316--338~$\mu$s and approximately 40~Mbps goodput. At $J=2$, four of five completed deployments meet all criteria, delivering 79.96--79.99~Mbps. UL p99 spans 360--515~$\mu$s, with up to 1.97\% of active intervals above budget; receive timeouts reach 48.8\% of all intervals in one cell. Completed deployments can still contain timing or synchronization failures.

At $J=4$, goodput is 159.09--159.82~Mbps and DL p99 is 294--363~$\mu$s. The 19 plotted UL p99 values span 439--564~$\mu$s. Active UL intervals above budget range from 0.14\% to 5.40\%, and two completed deployments include resynchronization. Every completed cell maintains real-time radio-clock progression, but no four-cell deployment meets all criteria. Thus, two-cell operation meets the criteria in four completed runs, while only the single-cell configuration meets them in all five.

\subsection{OCUDU SISO Validation}
\label{subsec:eval_ocudu}

We validate OCUDU operation with OAI software UEs in a SISO configuration using a 3.75~GHz carrier, 40~MHz bandwidth, 30~kHz SCS, and 61.44~Msps sampling. With one UE and an ideal channel, the UE completes synchronization, random access, registration, and IP connectivity. A 10~s traffic test delivers receiver throughputs of 12.05~Mbps downlink and 11.91~Mbps uplink. These measurements confirm end-to-end operation of a second gNB stack through the same \mirage IQ interface.

\subsection{Scope and Limitations}
The H100 evaluation supports the tested filtering accuracy, update consistency, and workload-dependent acceleration. It does not establish saturation capacity, rare-failure reliability, full-stack delivery latency, or long-term reliability of concurrent cells. The observed bidirectional synchronization failures limit full-stack reliability. The reported processing times apply to the tests described in Section~\ref{subsec:eval_h100}.

Our implementation copies IQ samples between host and GPU memory, including when the host buffers are registered with CUDA. Grace Hopper provides coherent CPU--GPU memory access through NVLink-C2C~\cite{nvidia_grace_hopper_architecture}, which could allow the emulator to access CPU-resident IQ buffers without explicit copies. Whether this improves processing latency or reduces deadline misses requires measurement and depends on memory placement, access patterns, and synchronization.

The separate physical-network case study is presented in Section~\ref{sec:dt}.

\section{Digital Twin-Assisted Network Control}
\label{sec:dt}

We evaluate \mirage in a slice-control experiment coupled to the X5G physical \gls{ran}. The case study compares physical and twin throughput under \gls{los} and \gls{nlos} conditions, then examines the transfer of PRB quotas selected in the twin to the physical gNB. It assesses how prediction errors affect the number of physical reconfigurations required to meet a throughput cap.

\begin{figure}
    \centering
    \resizebox{\linewidth}{!}{\input{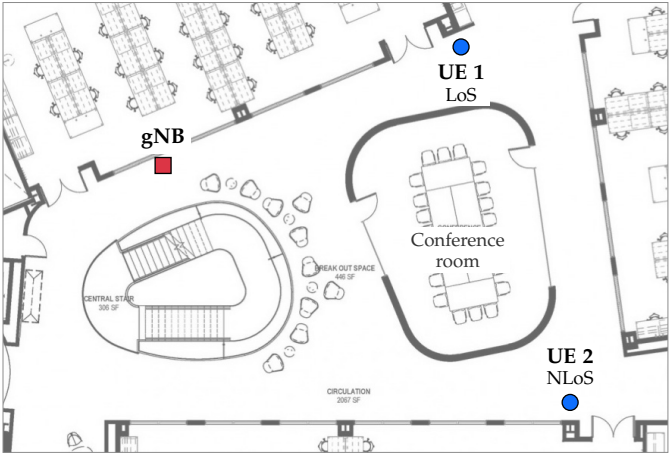}}
    \caption{Deployment on the sixth floor of Northeastern University's EXP building. UE~1 has a \gls{los} path to the gNB; the conference room obstructs the direct path to UE~2 (\gls{nlos}).}
    \label{fig:floor}
\end{figure}

\subsection{Scenario and Testbed Configuration}
\label{subsec:dt_setup}

The experiment uses the indoor X5G deployment on the sixth floor of Northeastern University's EXP building (Fig.~\ref{fig:floor}). UE~1 has a \gls{los} path to the gNB, whereas a conference room obstructs the direct path to UE~2. Each UE belongs to a separate slice: \gls{sd}~0 for UE~1 and SD~1 for UE~2.

The physical gNB uses a Foxconn RPQN O-RU at $3.75$~GHz with $40$~MHz bandwidth, $30$~kHz \gls{scs}, $2\times2$ antennas, and TDD pattern \texttt{DDDDDDDSUU}. It runs an \gls{oai} \gls{du}/\gls{cu} stack with an Open5GS \gls{5gc}; the UEs are Samsung Galaxy S23 handsets. The twin uses the corresponding OAI and Open5GS configuration, with OAI \texttt{nr-uesoftmodem} processes replacing the handsets and the \texttt{gpuemu} driver connecting the stacks to \mirage. Numerology, bandwidth, TDD pattern, MCS table, slice configuration, and antenna count are matched between the two domains. Receiver implementation and hardware differences remain possible sources of throughput disagreement.

Sionna~RT uses an imported CAD model of the floor and gNB and UE coordinates measured against the floor plan. It computes per-UE \glspl{cfr} $\mathbf{H}_k[q]$ on the emulator's FFT grid and publishes them through Redis at approximately $10$~Hz. This quasi-static scenario uses configured geometry and positions rather than continuous environmental sensing. The monitoring and control xApps described in Section~\ref{sec:design} collect per-UE throughput and apply slice-level PRB quotas through E2.

\subsection{Network Policy and Closed-Loop Procedure}
\label{subsec:dt_loop}

Let $s\in\{0,1\}$ denote the slice index, with UE~$k=s+1$ assigned to slice~$s$. The policy imposes a downlink throughput cap:
\begin{equation}
    T_s \leq T_s^{\max}, \qquad s\in\{0,1\},
    \label{eq:thr_threshold}
\end{equation}
where $T_s$ is measured slice throughput. The caps are $T_0^{\max}=25$~Mbps and $T_1^{\max}=10$~Mbps. The controller adjusts each slice's maximum PRB quota $\rho_s$. These quotas bound resource use; they do not specify a fixed partition of all PRBs. Because throughput also depends on channel and receiver behavior, a quota satisfying the cap in the twin may not satisfy it on the physical link.

Candidate quotas are evaluated sequentially in one twin. Let $\tau_\mathrm{warm}$ denote the baseline warm-up duration and $\tau_\mathrm{meas}$ the measurement duration for each candidate and its physical validation, both in seconds. The procedure has four phases:

\smallskip\noindent\textbf{Phase~1 -- Baseline monitoring.} The physical deployment and the twin run with the scheduler's default allocation for $\tau_\mathrm{warm}$ seconds. The \gls{kpm} xApp records per-slice throughput $T_s^{\mathrm{phy}}$ and $\hat T_s^{\mathrm{DT}}$ in Redis.

\smallskip\noindent\textbf{Phase~2 -- Violation detection.} A twin throughput measurement exceeding $T_s^{\max}$ triggers quota exploration for slice~$s$ through the broker.

\smallskip\noindent\textbf{Phase~3 -- Quota exploration.} The RC xApp reduces the violating slice's maximum quota from $100\%$ in steps of $\Delta\rho=5$ percentage points, within the range $[10\%,100\%]$. Each candidate is applied to the twin and evaluated over $\tau_\mathrm{meas}$ seconds. The search selects the first candidate meeting~\eqref{eq:thr_threshold}, denoted $\rho_s^\star$. This descending search seeks a less restrictive quota, but time-varying measurements do not establish a globally optimal allocation.

\smallskip\noindent\textbf{Phase~4 -- Deployment and monitoring.} The RC xApp applies $\rho_s^\star$ to the physical slice, whose throughput is recorded for $\tau_\mathrm{meas}$ seconds. Twin monitoring continues after deployment. If the twin again exceeds the cap, exploration resumes from the current quota and may produce another physical update. Thus, renewed exploration is triggered by the twin measurements; a physical violation alone does not establish that another search will occur.

\subsection{Digital Twin Fidelity Assessment}
\label{subsec:dt_fidelity}

Before control, each UE is offered a $50$~Mbps downlink \gls{udp} flow generated using \texttt{iperf3}, with the core-side endpoint co-located with the \gls{upf}. The \gls{kpm} xApp collects \texttt{DRB.UEThpDl} from both gNBs. For slice~$s$, let $\bar T_s^{\mathrm{phy}}$ and $\overline{\hat T}_s^{\mathrm{DT}}$ denote mean throughput over the steady-state comparison window. The signed relative error is
\begin{equation}
    \epsilon_s = \frac{\bar T_s^{\mathrm{phy}}-\overline{\hat T}_s^{\mathrm{DT}}}{\bar T_s^{\mathrm{phy}}}, \qquad s\in\{0,1\}.
    \label{eq:throughput_error}
\end{equation}
Positive values indicate underestimation by the twin; negative values indicate overestimation.

\begin{figure}
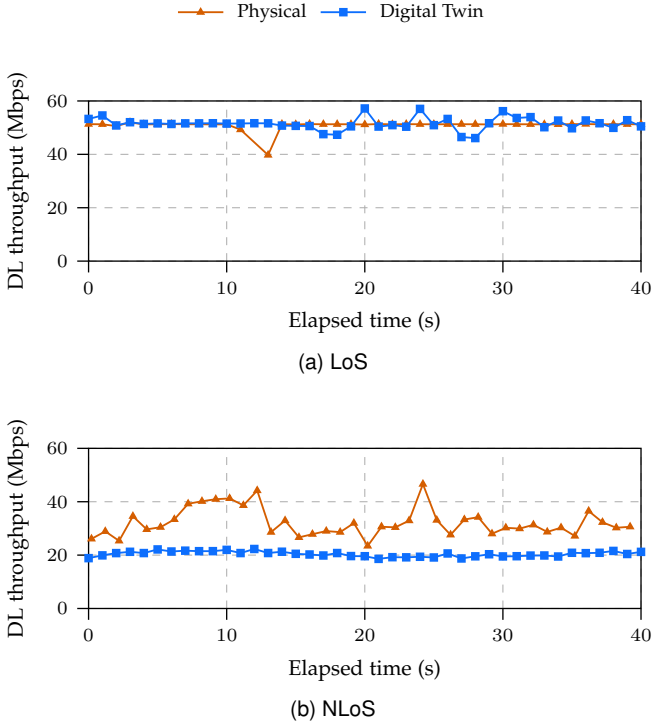

    \centering
    \input{figs/tikz/dt_plot_style.tex}
% Shared legend for the paired panels.
\begin{tikzpicture}
\begin{axis}[hide axis,scale only axis,width=1mm,height=1mm,
xmin=0,xmax=1,ymin=0,ymax=1,
legend columns=2,legend cell align=left,
legend style={font=\scriptsize,draw=none,inner sep=1pt,column sep=3pt,
at={(0,0)},anchor=center}]
\addlegendimage{color=physicalWarm,mark=triangle*,mark size=1.25}
\addlegendentry{Physical}
\addlegendimage{color=dtBlue,mark=square*,mark size=1.25}
\addlegendentry{Digital Twin}
\end{axis}
\end{tikzpicture}\par\smallskip
    \setlength{\tabcolsep}{0pt}
    \subfloat[LoS\label{fig:los_dl}]{\begin{minipage}[b]{\linewidth}
        \centering
        \input{figs/tikz/dt_plot_style.tex}
% This file was created with tikzplotlib v0.10.1.
\begin{tikzpicture}

\definecolor{darkgray176}{RGB}{176,176,176}
\definecolor{lightgray204}{RGB}{204,204,204}

\begin{axis}[
height=3.7cm,
legend cell align={left},
legend columns=2,
legend style={
  fill opacity=0.8,
  draw opacity=1,
  text opacity=1,
  at={(0.5,1.02)},
  anchor=south,
  draw=lightgray204
},
tick align=outside,
tick pos=left,
width=\linewidth,
x grid style={darkgray176},
xlabel={Elapsed time (s)},
xmajorgrids,
xmin=0, xmax=40,
xtick style={color=black},
y grid style={darkgray176},
ylabel={DL throughput (Mbps)},
ymajorgrids,
ymin=0, ymax=60,
ytick style={color=black}
]
\addplot [physicalWarm, mark=triangle*, mark size=1.25, mark options={solid, fill opacity=1, draw opacity=1}]
table {%
0 51.319664
1.001 51.266528
2 50.592008
3 51.971784
4 51.222432
5 51.266584
6 51.342152
7 51.27152
8 51.27236
9 51.219128
10.001 51.238672
11 49.276352
13 39.740224
14 51.311744
15.001 51.203336
16 51.336888
17.001 51.272152
18 51.27984
19 51.268376
20 51.168128
21 51.359304
22 51.239472
23 51.315672
24 51.282928
25 51.267496
26 51.27792
27 51.2914
28 51.270112
29 51.20144
30 51.324368
31 51.277144
32 51.269448
33 51.222664
34 51.23908
35 51.304424
36 51.346272
37 51.268136
38 51.291096
39 51.27172
40 51.267368
};

\addplot [dtBlue, mark=square*, mark size=1.25, mark options={solid, fill opacity=1, draw opacity=1}]
table {%
0.000999999999999446 53.26884
1 54.564112
2 50.823352
3 52.039232
4 51.39644
5 51.595928
6 51.381952
7 51.612216
8 51.607032
9 51.652104
10 51.519864
11.004 51.561128
12 51.654648
13 51.624024
14 50.774496
15 50.673304
16 50.604304
17 47.570664
18 47.346272
19 50.483424
20 57.190744
21 50.405552
22 50.94812
23 50.35712
24 57.033072
25 50.9356
26 53.26528
27 46.482736
28 46.098528
29 51.657448
30 56.170528
31 53.624608
32 53.942128
33 50.167456
34 52.662304
35 49.772328
36 52.692672
37 51.65684
38 49.931672
39 52.791128
40 50.482112
};

\end{axis}

\end{tikzpicture}

    \end{minipage}}\par\smallskip
    \subfloat[NLoS\label{fig:nlos_dl}]{\begin{minipage}[b]{\linewidth}
        \centering
        \input{figs/tikz/dt_plot_style.tex}
% This file was created with tikzplotlib v0.10.1.
\begin{tikzpicture}

\definecolor{darkgray176}{RGB}{176,176,176}
\definecolor{lightgray204}{RGB}{204,204,204}

\begin{axis}[
height=3.7cm,
legend cell align={left},
legend columns=2,
legend style={
  fill opacity=0.8,
  draw opacity=1,
  text opacity=1,
  at={(0.5,1.02)},
  anchor=south,
  draw=lightgray204
},
tick align=outside,
tick pos=left,
width=\linewidth,
x grid style={darkgray176},
xlabel={Elapsed time (s)},
xmajorgrids,
xmin=0, xmax=40,
xtick style={color=black},
y grid style={darkgray176},
ylabel={DL throughput (Mbps)},
ymajorgrids,
ymin=0, ymax=60,
ytick style={color=black}
]
\addplot [physicalWarm, mark=triangle*, mark size=1.25, mark options={solid, fill opacity=1, draw opacity=1}]
table {%
0.210000000000001 26.083856
1.21 28.894616
2.21 25.341952
3.21 34.559352
4.21 29.636256
5.21 30.456584
6.21 33.378488
7.21 39.236048
8.21 40.175768
9.21 40.94248
10.212 41.2406
11.21 38.633112
12.211 44.197472
13.21 28.599616
14.229 32.956832
15.233 26.643528
16.21 27.83508
17.21 29.00952
18.217 28.608192
19.21 31.98284
20.21 23.420432
21.21 30.624192
22.21 30.40132
23.21 32.928616
24.21 46.60256
25.21 33.116216
26.21 27.596384
27.21 33.32288
28.21 34.221864
29.21 28.00688
30.21 30.2366
31.21 29.928744
32.21 31.350688
33.21 28.685608
34.21 30.285992
35.21 27.152992
36.21 36.51416
37.21 32.301024
38.21 30.24828
39.21 30.617568
};

\addplot [dtBlue, mark=square*, mark size=1.25, mark options={solid, fill opacity=1, draw opacity=1}]
table {%
0.000999999999999446 18.865928
1 19.903152
2 20.751912
3 21.241504
4 20.765016
5 22.12836
6 21.348672
7 21.661472
8 21.474408
9 21.483376
10 21.966344
11.004 20.7674
12 22.2946
13 20.780808
14 21.270112
15.001 20.487168
16 20.23012
17 19.857088
18 20.777664
19 19.639088
20 19.588912
21 18.591216
22 19.224208
23 19.170336
24 19.378216
25 19.118848
26 20.62236
27 18.74292
28.001 19.54776
29 20.325824
30 19.504768
31 19.586712
32 19.825744
33 19.86672
34 19.475768
35.001 20.8732
36 20.734168
37.001 20.86944
38 21.570592
39 20.416072
40 21.215696
};

\end{axis}

\end{tikzpicture}

    \end{minipage}}
    \caption{Per-UE downlink throughput (\texttt{DRB.UEThpDl}) on the physical deployment and the digital twin under identical traffic load. SD\,=\,0: LoS UE; SD\,=\,1: NLoS UE.}
    \label{fig:kpm_comparison}
\end{figure}

Figure~\ref{fig:kpm_comparison} shows the throughput traces. The mean errors are $\epsilon_0=-1.45\%$ for the \gls{los} UE and $\epsilon_1=36.5\%$ for the \gls{nlos} UE. The LoS means agree closely under the tested offered load. Since both are near $50$~Mbps, this agreement does not establish equal link capacity or channel fidelity. The NLoS twin underestimates physical throughput and exhibits less variability in the displayed trace. Propagation modeling, receiver implementation, and hardware differences may contribute to this discrepancy; the experiment does not isolate their effects. The control experiment below examines the consequence of this disagreement for quota selection.

\subsection{Closed-Loop Control Results}
\label{subsec:cl_results}

\begin{figure}
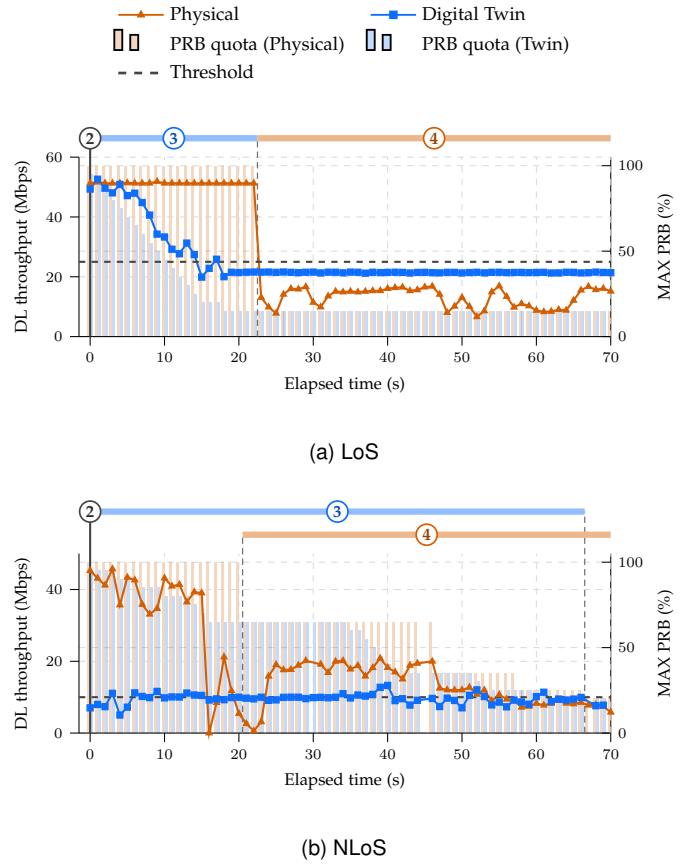
\centering
  \input{figs/tikz/dt_plot_style.tex}
% Standalone shared legend. Preamble: \usepackage{pgfplots}.
\begin{tikzpicture}
\colorlet{physical}{physicalWarm}
\colorlet{digital}{dtBlue}
\colorlet{thresholdc}{policyThreshold}
\definecolor{darkgray176}{RGB}{176,176,176}
\definecolor{lightgray204}{RGB}{204,204,204}
\begin{axis}[
hide axis,
scale only axis,
width=1mm,
height=1mm,
xmin=0,
xmax=1,
ymin=0,
ymax=1,
legend columns=2,
legend cell align={left},
legend style={font=\scriptsize,draw=none, at={(0,0)}, anchor=center, /tikz/every even column/.append style={column sep=6pt}}
]
\addlegendimage{physical, mark=triangle*, mark size=1.25, mark options={fill opacity=1, draw opacity=1}}
\addlegendentry{Physical}
\addlegendimage{digital, mark=square*, mark size=1.25, mark options={fill opacity=1, draw opacity=1}}
\addlegendentry{Digital Twin}
\addlegendimage{ybar, ybar legend, draw=none, fill=physical, fill opacity=0.25}
\addlegendentry{PRB quota (Physical)}
\addlegendimage{ybar, ybar legend, draw=none, fill=digital, fill opacity=0.25}
\addlegendentry{PRB quota (Twin)}
\addlegendimage{thresholdc, dashed, line width=1pt}
\addlegendentry{Threshold}
\end{axis}
\end{tikzpicture}\par\smallskip
  \subfloat[LoS\label{fig:kpm_control_los}]{\begin{minipage}[b]{\linewidth}
  \input{figs/tikz/kpm_control_los}

  \end{minipage}}\par\smallskip
  \subfloat[NLoS\label{fig:kpm_control_nlos}]{\begin{minipage}[b]{\linewidth}
  \input{figs/tikz/kpm_control_nlos}

  \end{minipage}}
  \caption{Slice throughput and maximum PRB quotas during closed-loop control. Lines represent throughput, bars indicate PRB quotas, and horizontal lines mark the throughput caps of 25~Mbps for LoS and 10~Mbps for NLoS conditions. The closed-loop control phases are indicated at the top of the graph. Phase 1 is omitted because the graph begins with an already-occurring threshold violation.}
\end{figure}

For the \gls{los} slice, the twin initially reports approximately $50$~Mbps, exceeding the $25$~Mbps cap. The search evaluates 16 intermediate quotas before selecting $\rho_0^\star=15\%$. The physical gNB receives one update, and its throughput falls below the cap at approximately $23$~s (Fig.~\ref{fig:kpm_control_los}). After the quota is applied, the physical and twin throughputs differ, although both satisfy the cap. Thus, the close baseline means do not imply equally close predictions under a restricted quota.

\begin{figure}
    \centering
    \input{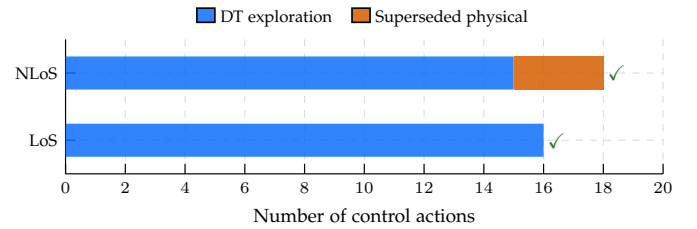}
\resizebox{\linewidth}{!}{%
\begin{tikzpicture}
\colorlet{ddt}{dtBlue}
\definecolor{okgreen}{HTML}{2E7D32}
\colorlet{dphys}{physicalWarm}
\definecolor{darkgray176}{RGB}{176,176,176}
\definecolor{lightgray204}{RGB}{204,204,204}
\pgfplotsset{every axis legend/.append style={draw=none, fill=none}}
\begin{axis}[
xbar stacked,
width=\linewidth,
height=2cm,
scale only axis,
bar width=14pt,
enlarge y limits=0.5,
symbolic y coords={LoS,NLoS},
ytick=data,
xmin=0,
xmax=20,
xlabel={Number of control actions},
axis x line*=bottom,
axis y line*=left,
xmajorgrids,
grid style={darkgray176, line width=0.3pt},
xtick style={color=black},
ytick style={color=black},
legend cell align={left},
legend columns=-1,
legend style={
  draw=none,
  fill=none,
  cells={anchor=west},
  at={(0.5,1.03)},
  anchor=south,
  /tikz/every even column/.append style={column sep=6pt}
}
]
\addplot[xbar, draw=none, fill=ddt, fill opacity=0.85] coordinates {(16,LoS) (15,NLoS)};
\addlegendentry{DT exploration}
\addplot[xbar, draw=none, fill=dphys, fill opacity=0.85] coordinates {(0,LoS) (3,NLoS)};
\addlegendentry{Superseded physical}
% green check marks the converged (final, compliant) control action on each slice
\node[anchor=west, inner sep=1.5pt, okgreen, font=\small] at (axis cs:16,LoS) {$\checkmark$};
\node[anchor=west, inner sep=1.5pt, okgreen, font=\small] at (axis cs:18,NLoS) {$\checkmark$};
\end{axis}
\end{tikzpicture}%
}
    \caption{Intermediate actions before the final selected quota. DT exploration counts intermediate candidate quotas tested in the twin; superseded physical counts deployed quotas later replaced. The check mark denotes the final physical action, which is excluded from the bars.}
    \label{fig:control_exploration}

\end{figure}

For the \gls{nlos} slice, twin throughput fluctuates near the $10$~Mbps cap while physical throughput is higher during much of the run (Fig.~\ref{fig:kpm_control_nlos}). The controller successively deploys quotas of $65\%$, $35\%$, $25\%$, and $20\%$. Twin threshold crossings lead to resumed exploration and further quota reductions. Figure~\ref{fig:control_exploration} records 15 intermediate twin candidates and three superseded physical updates, excluding the final action. Physical throughput is below the cap by the final update, but it is also below the cap during the preceding $25\%$ interval. The trace therefore supports eventual compliance without establishing that the final $20\%$ quota is the least restrictive physical setting. Throughput changes before and between updates also prevent attributing every change to a quota adjustment alone.

In the observed control traces, both slices ultimately satisfy their throughput caps. The LoS case requires one physical reconfiguration, whereas the NLoS case requires four. This contrast illustrates how a twin can confine candidate testing to the digital environment when its predictions support the control decision, while prediction errors near the threshold can lead to repeated physical reconfiguration. The result does not establish guaranteed recovery from a physical violation: further exploration depends on another twin-side trigger. Improved prediction accuracy could reduce such reconfigurations, but that benefit requires validation under repeated trials and additional channel conditions.

\section{Conclusions}
\label{sec:conclusion}

We presented \mirage, a GPU-accelerated, IQ-level channel emulator connected to OAI and OCUDU 5G NR stacks through radio-driver plugins. The design combines batched SISO/MIMO frequency-domain filtering, timestamped shared-memory transport, and versioned channel responses. External propagation models supply channel updates asynchronously, while IQ processing continues with a complete coefficient version for each link and direction.

Component tests on an NVIDIA H100 confirm filtering accuracy and channel-update consistency under the tested conditions. At $61.44$~Msps, the highest tested configurations meeting the bidirectional $0.5$~ms processing criterion are eight SISO links, four $2\times2$ links, and one $4\times4$ link. Matched CPU comparisons show that GPU acceleration depends on workload size. OAI full-stack tests demonstrate traffic delivery at the evaluated loads, but synchronization failures limit reliability. OCUDU SISO tests also confirm end-to-end operation with OAI UEs using the same emulator. With dedicated MIG partitions, completed deployments maintain real-time radio-clock progression up to four cells. All criteria are met in five of five completed single-cell runs and four of five two-cell runs, but in none of the four-cell runs.

The X5G case study uses the emulator to evaluate candidate slice quotas in the twin before applying selected settings to the physical gNB. Both slices ultimately meet their throughput caps in the observed traces, with one physical reconfiguration in the LoS case and four in the NLoS case. These results highlight the importance of prediction accuracy near the control threshold when applying decisions from the twin to the physical network. They motivate further work on full-stack synchronization, resource scheduling across concurrent instances, and calibration against physical measurements.

\balance
%% References
\bibliographystyle{IEEEtran}
%\bibliography{IEEEabrv,./references}
\bibliography{biblio.bib}

\end{document}